%% file: Main.tex
\documentclass[12pt]{article}
\pdfoutput=1

\input{Preamble}

\begin{document}

\input{Titlepage.tex}

\input{Introduction.tex}
\input{Sections.tex}
\input{AllTops.tex}
\input{U1charges.tex}
\input{Flatness.tex}

\input{Conclusions.tex}

\section*{Acknowledgments}

We would like to thank Philip Candelas, Denis Klevers, Eran Palti,
Raffaele Savelli, Sakura Sch\"afer-Nameki, and Timo Weigand for
interesting discussions.  The work of T.G.~and J.K.~is supported by a
grant of the Max Planck Society.

\appendix
\input{EllipticCurves.tex}

\bibliography{Volker,Jan,Thomas}
\bibliographystyle{utcaps}

\end{document}

%% file: Preamble.tex
\newcommand{\VersionInformation}{}  
\InputIfFileExists{Debug}{}{}

\usepackage[utf8]{inputenc}
\usepackage{amsmath}
\usepackage{amsthm}
\usepackage{amsfonts}          
\usepackage{amssymb}           
\usepackage[mathscr]{eucal}
\usepackage{graphicx}
\usepackage{color}
\usepackage{hypbmsec}
\usepackage{fancyvrb}
\usepackage{sepnum}
\usepackage{xspace}
\usepackage{booktabs}
\usepackage{rotating}
\usepackage{multirow}
\usepackage[vcentermath]{youngtab}
\usepackage{verbatim}
\usepackage[isu,bf]{caption}
\usepackage{array}

\newlength{\xtrawidth}
\newlength{\xtraheight}
\usepackage[all]{xy}           

\ifx\NoBackreferences\EMPTYMACRO
  \usepackage[backref,linktocpage,bookmarks]{hyperref}
\else
  \usepackage[linktocpage,bookmarks]{hyperref}
\fi

\ifx\DEBUG\EMPTYMACRO
\else
  \usepackage{showlabels}
\fi

\def\clap#1{\hbox to 0pt{\hss#1\hss}}

\makeatletter
  \def\adots{\mathinner{\mkern2mu\raise\p@\hbox{.}
      \mkern2mu\raise4\p@\hbox{.}\mkern1mu
      \raise7\p@\vbox{\kern7\p@\hbox{.}}\mkern1mu}}
\makeatother

\newcommand{\eqdef}{%
  \mathrel{\lower.1mm
    \hbox{$\stackrel{\lower.424ex\hbox{\scriptsize def}}{=}$}}
}
\newcommand{\Q}{\ensuremath{{\mathbb{Q}}}}
\newcommand{\R}{\ensuremath{{\mathbb{R}}}}
\newcommand{\C}{\ensuremath{{\mathbb{C}}}}

\newcommand{\Z}{\mathbb{Z}}
\newcommand{\CP}{{\ensuremath{\mathop{\null {\mathbb{P}}}\nolimits}}}

\newcommand{\ptset}{\ensuremath{\{\text{pt.}\}}}

\DeclareMathOperator{\MW}{MW}

\DeclareMathOperator{\Span}{span}

\DeclareMathOperator{\rank}{rank}

\newcommand{\Rep}[1]{\ensuremath{\mathbf{#1}}}

\newcommand{\tmod}{~\mathrm{mod}~}

\DeclareMathOperator{\conv}{conv}

{\end{list}}

{\everymath{\displaystyle\everymath{}}\array}%
{\endarray}

\usepackage{color}
\usepackage{listings} 
\usepackage{courier}

\definecolor{dbluecolor}{rgb}{0.01,0.02,0.7}
\definecolor{dgreencolor}{rgb}{0.2,0.4,0.0}
\definecolor{dgraycolor}{rgb}{0.30,0.3,0.30}

\lstdefinelanguage{SageInputLanguage}{
  language=Python, 
  morekeywords={False,sage,True},
  sensitive=true,
}

\lstdefinestyle{SageInput}{
  language=SageInputLanguage,
  basicstyle=\fontsize{11pt}{11pt}\ttfamily\bfseries,
  commentstyle={\ttfamily\color{dgreencolor}},
  stringstyle={\color{dgraycolor}\bfseries},
  keywordstyle=\ttfamily\color{dbluecolor}\bfseries\color{red},
  xleftmargin=25pt,
  belowskip=3pt,
}

\lstdefinelanguage{SageOutputLanguage}{
  morekeywords={False,True},
  sensitive=true,
}

\lstdefinestyle{SageOutput}{
  language=SageOutputLanguage,
  basicstyle={\fontsize{11pt}{11pt}\ttfamily},
  commentstyle={\ttfamily\color{dgreencolor}},
  keywordstyle={\ttfamily\color{dbluecolor}},
  stringstyle={\ttfamily\color{dgraycolor}},
  xleftmargin=25pt,
  aboveskip=0pt,
}

\lstdefinestyle{DefaultSageInputOutput}{
  identifierstyle=,
  numbersep=5pt,
  aboveskip=0pt,
  belowskip=0pt,
  breaklines=true,
  numberstyle=\footnotesize,
  numbers=right,
}


%% file: Titlepage.tex
\begin{titlepage}
\begin{flushright}
\parbox[t]{1.8in}{\begin{flushright} MPP-2013-144 \end{flushright}}
\end{flushright}

\begin{center}

\vspace*{ 1.2cm}


\textbf{\LARGE\boldmath 
  Geometric Engineering in Toric F-Theory\\[4pt]
  and GUTs with $U(1)$ Gauge Factors
}

\vskip 1cm

\renewcommand{\thefootnote}{}
\begin{center}
  Volker Braun,${}^1$ 
  Thomas W.~Grimm,${}^2$
  and Jan Keitel${}^2$ 
  \footnote{\texttt{volker.braun@maths.ox.ac.uk}; \texttt{grimm}, \texttt{jkeitel@mpp.mpg.de}}
\end{center}
\vskip .2cm
\renewcommand{\thefootnote}{\arabic{footnote}}

{${}^1$Mathematical Institute,
  University of Oxford\\
  24-29 St Giles', Oxford, OX1 3LB, United Kingdom}

\vspace*{.2cm}
 
{${}^2$Max-Planck-Institut f\"ur Physik, \\
F\"ohringer Ring 6, 80805 Munich, Germany}

 \vspace*{.8cm}

\end{center}

\vskip 0.2cm
 
\begin{center} {\bf ABSTRACT } \end{center}

An algorithm to systematically construct all Calabi-Yau elliptic fibrations  
realized as hypersurfaces in a toric ambient space
for a given base and gauge group is described. 
This general method is applied to the particular question of
constructing $SU(5)$ GUTs with multiple $U(1)$ gauge factors. 
The basic data consists of a \emph{top} over each toric divisor in the base together
with compactification data giving the embedding into a reflexive
polytope. The allowed choices of compactification data are integral
points in an auxiliary polytope. In order to ensure the existence of a
low-energy gauge theory, the elliptic fibration must be flat, which is 
reformulated into conditions on the top and its embedding. In
particular, flatness of $SU(5)$ fourfolds imposes additional linear
constraints on the auxiliary polytope of compactifications, and is
therefore non-generic.
Abelian gauge symmetries arising in toric F-theory compactifications
are studied systematically. Associated to each top, the toric
Mordell-Weil group determining the minimal number of $U(1)$ factors is
computed. Furthermore, all $SU(5)$-tops and their splitting types
are determined and used to infer the pattern of $U(1)$ matter charges.

\vspace*{.5cm}

\hfill {June 3, 2013}
\end{titlepage}

\VersionInformation

\tableofcontents
\listoffigures
\listoftables


%% file: Introduction.tex
\section{Introduction}
\label{sec:intro}

\begin{figure}[htb]
  \centering
  \includegraphics[width=\textwidth]{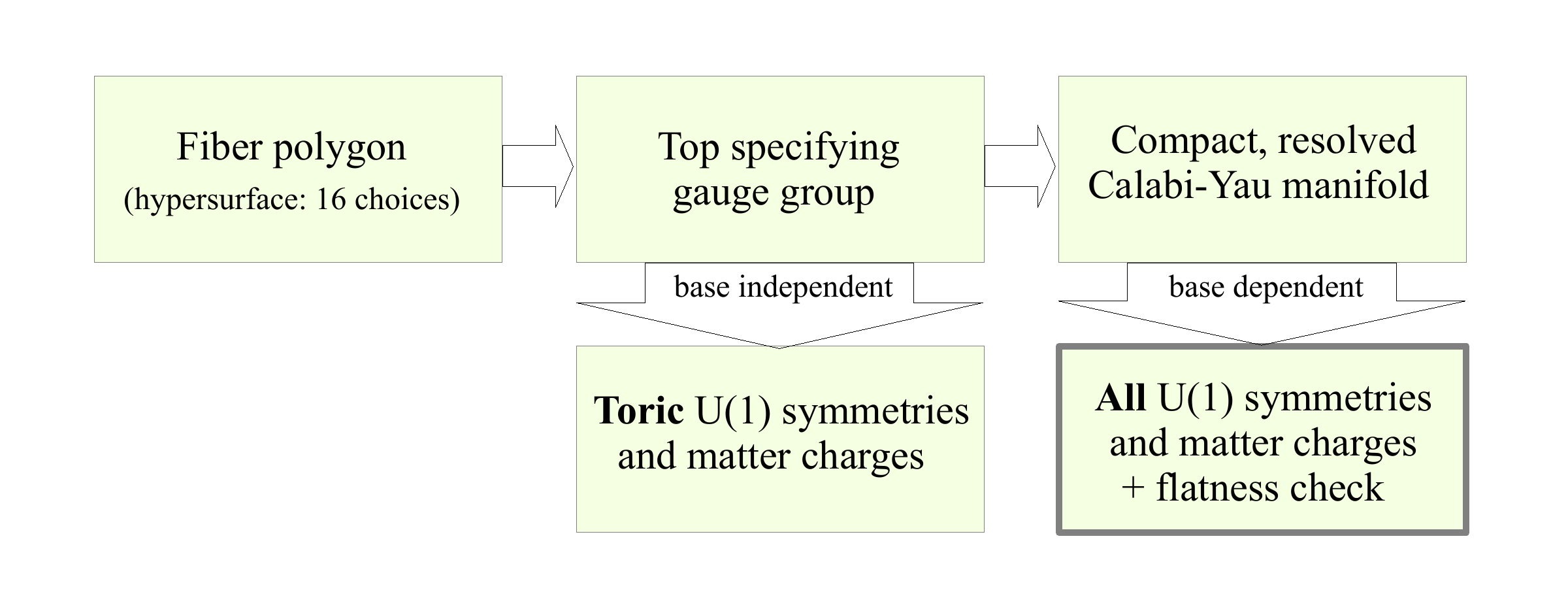}
  \caption[Toric F-theory flowchart]{Systematical approach to constructing
    compact F-theory backgrounds with specified gauge group and base
    space.}
  \label{fig:flow_chart}
\end{figure}

Four-dimensional F-theory compactifications on elliptically fibered
Calabi-Yau fourfolds allow for the exciting possibility to
geometrically engineer interesting Grand Unified Theories
(GUTs)~\cite{Donagi:2008ca, Beasley:2008dc}.  While the original
models were local, and thus decoupled from gravity, vast progress has
been made in the search for global realizations~\cite{Marsano:2009gv,
  Blumenhagen:2009yv, Marsano:2009wr, Grimm:2009yu, Chen:2010ts,
  Knapp:2011wk, Weigand:2010wm, Maharana:2012tu}. To systematically
approach this problem, one has to construct fully resolvable
Calabi-Yau fourfolds, e.g.~with $SU(5)$ and $SO(10)$ gauge group
singularities~\cite{Blumenhagen:2009yv, Grimm:2009yu, Chen:2010ts,
  Knapp:2011wk}, but also needs to globally understand the physics of
Abelian gauge symmetries. Abelian gauge symmetries are not localized
on the degenerate fibers and, hence, depend on the global properties
of the Calabi-Yau manifold. Therefore, a systematic study of allowed
models requires us to control both the local singularity resolution
for the non-Abelian groups and the global properties of $U(1)$ gauge
symmetries. Studying global $U(1)$ symmetries is a topic of intense
current research~\cite{Grimm:2010ez, Park:2011ji, Morrison:2012ei,
  Cvetic:2012xn, Mayrhofer:2012zy, Braun:2013yti, Borchmann:2013jwa,
  Cvetic:2013nia, Grimm:2013oga}. In this paper, we use toric geometry
to develop an algorithm to construct compact Calabi-Yau fourfolds with
a certain non-Abelian and Abelian gauge group for a specified base.
The various steps are schematically pictured in
\autoref{fig:flow_chart} and we will discuss them in the following.

In F-theory, the physics of space-time filling seven-branes is encoded
in the degeneration of the elliptic fiber and the sections of the
fibration. First, one has to specify a zero-section that embeds the
base used as a Type IIB string background into the Calabi-Yau
manifold.  In this work we will only consider elliptic fibers that are
hypersurfaces inside a toric ambient space. This leaves $16$ choices
for the fiber ambient space corresponding to the $16$ inequivalent
reflexive polygons in two dimensions. For a given fiber ambient
spaces, one can then classify all \emph{toric} sections in the
corresponding elliptically fibered manifold. These form the toric
Mordell-Weil group and determine the minimal number of $U(1)$ factors
in the four-dimensional effective theory. Because the fiber ambient
space is the same for almost all fibers, the number of toric sections
is independent of the chosen base and non-Abelian gauge group. As a
first step, we will classify all such toric sections, their
dependencies, and the toric Mordell-Weil group they generate. While
this gives us a first understanding of allowed $U(1)$ symmetries, a
complete model requires one to also specify the non-Abelian gauge
symmetry and a base of the fibration, possibly giving rise to further
non-toric sections.

Next, to generate a non-Abelian gauge group, one modifies the toric
ambient space of the elliptic fiber over the seven-brane divisor in
the base. The Calabi-Yau hypersurfaces is then the resolution of the
non-Abelian singularity. The toric data is encoded in a so-called
\emph{top}~\cite{Candelas:1996su, Candelas:2012uu}. For each gauge
group there is a finite number of tops~\cite{Candelas:1996su,
  Bouchard:2003bu}. To actually classify all tops for a given gauge
group requires some care to mod out by the remaining discrete
symmetries of the top. As an example, we classify all $37$ distinct
$SU(5)$ tops for all $16$ fiber
types. Recently~\cite{Borchmann:2013jwa, Cvetic:2013nia}, one of the
fiber types together with its $5$ distinct $SU(5)$ tops was considered
in detail as a model for a rank-two Mordell-Weil group.

Fixing a particular top to obtain a desired gauge group determines the
toric $U(1)$ charges and constrains the $U(1)$ charges of the matter
fields. We demonstrate this for the example of $SU(5)$ with matter in
$\Rep{1}$, $\Rep{5}$ and $\Rep{10}$ representations. For each section
of the elliptic fibration, one can introduce a split of the affine
Dynkin diagram obtained in the resolution of the non-Abelian gauge
group.\footnote{In case of a single $U(1)$, our notion of
  \emph{splits} also captures the situations occurring in the
  split-spectral cover analysis performed in~\cite{Marsano:2009gv,
    Marsano:2009wr, Dudas:2010zb, Dolan:2011iu}.}  This split fixes
the $U(1)$ charges of the $\Rep{5}$, $\Rep{10}$ matter modulo
$5$. Moreover, a detailed analysis of the $\Rep{10}$ matter fields
shows that the fiber degeneration is fixed by the ambient space
fan. This forces them to all have the same $U(1)$ charge. Our no-go
result depends crucially on the fact that the fiber is a hypersurface of
codimension-one in the ambient space fiber, and can be avoided
using complete intersections~\cite{workinprogress}.

Finally, we present an algorithm to construct all reflexive polytopes
with a given top and base manifold. In the construction, we introduce
an auxiliary polytope whose lattice points label the resulting
inequivalent reflexive polytopes. Having found a method to construct
appropriate compactification manifolds, we further discuss flatness
which is tied to the compactification data. For phenomenological
reasons, one is generally interested in low-energy effective gauge
theories. Non-flat F-theory compactifications have tensionless
strings, yielding an infinite towers of massless fields in the
resulting effective theory. Therefore, the Calabi-Yau manifold must be
a \emph{flat} fibration, that is, have constant fiber dimension over
all points in the base. Ensuring flatness in various codimensions in
the base imposes additional constraints on the reflexive polytope
specifying the Calabi-Yau manifold and we formulate these in terms of
geometric conditions on the toric data.  Importantly, flatness is
non-generic in the sense that it corresponds to equations that points
of the auxiliary polytope have to satisfy. In fact, requiring the
fibration to be flat can rule out certain combinations of tops and
base manifold as well as entire tops independent of the base
manifold. To illustrate this point, we give examples for both cases.

The paper is organized as follows. In \autoref{sec:fiber}, we first
list the toric sections for all $16$ two-dimensional reflexive
polygons in which the elliptic fiber can be embedded. We then study
the toric subgroup of the Mordell-Weil group generated by them. In
\autoref{sec:tops}, all $SU(5)$ tops based on these $16$ fiber types
are listed and the number of induced toric $U(1)$ factors is
determined. These constrain the $U(1)$ matter charges for $SU(5)$
matter and singlets, which is discussed in
\autoref{sec:U(1)charges}. In particular, \emph{splits} of the tops
induced by the $U(1)$ factors are introduced and a no-go theorem for
$\Rep{10}$-matter with different $U(1)$ charges in hypersurfaces is
presented. In \autoref{sec:compactification}, we discuss the
compactification data necessary in addition to the choice of a tops
together with a base. Particular care has to be taken to ensure
flatness of the fibration, as we discuss in \autoref{sec:flatness}.
We introduce a toric flatness criterion and provide interesting
compact $SU(5)$ GUT examples in this final section.


%% file: Sections.tex
\section{Classifying Toric Sections}
\label{sec:fiber}

In order to classify the gauge theory content arising from F-theory
compactifications on Calabi-Yau manifolds, one might hope that
information about the gauge group is already specified by \emph{local}
properties of the geometry, as their classification could be
a more feasible task than trying to construct all possible Calabi-Yau
manifolds in various dimensions. For non-Abelian gauge groups, this does turn out to
be the case and the non-Abelian gauge group can be read off from the
local singularity structure of the compactification
manifold~\cite{Bershadsky:1996nh, Katz:1996th, Candelas:1996su, Candelas:1997eh,
Katz:2011qp}.

\begin{figure}
  \centering
  \includegraphics[width=\textwidth]{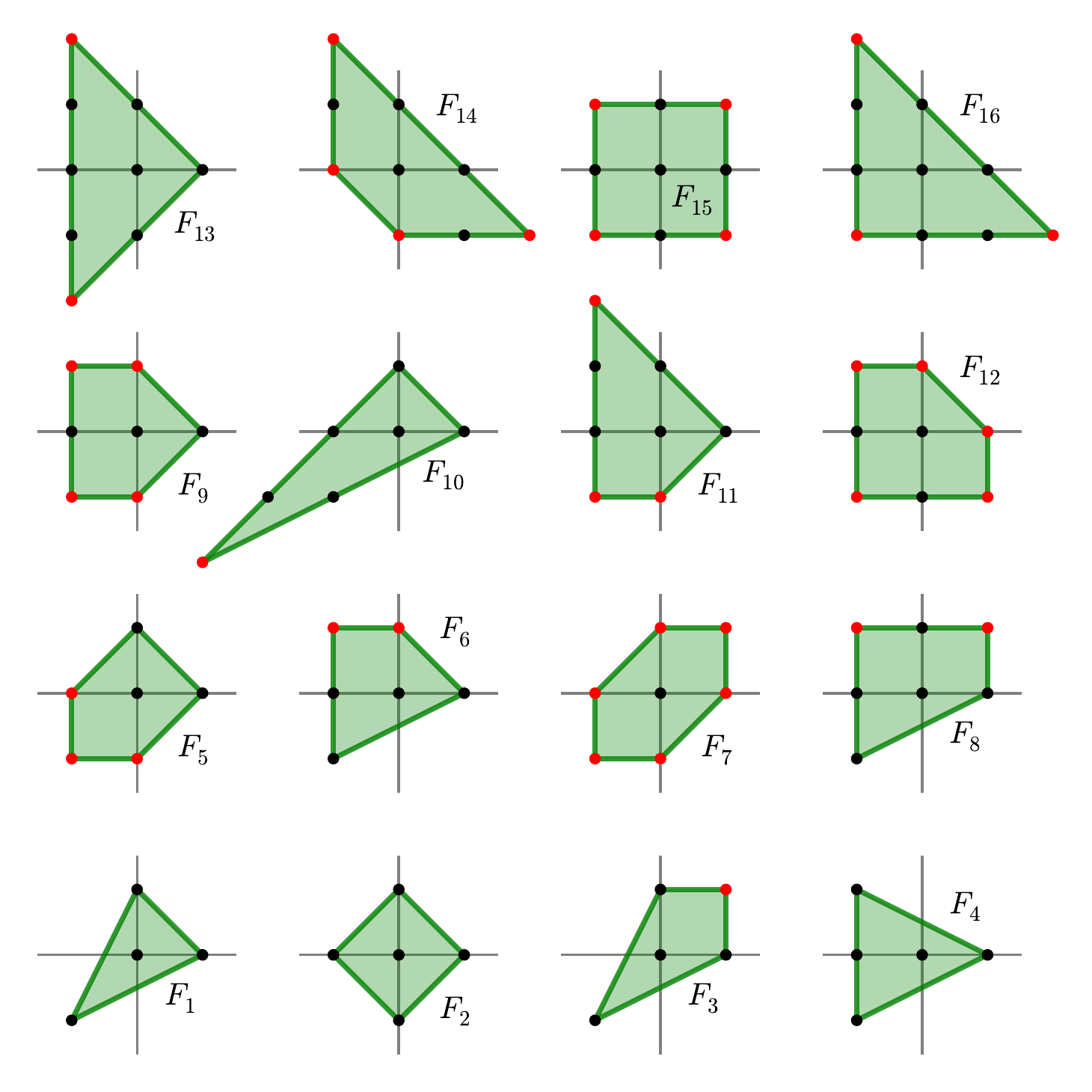}
  \caption[The $16$ reflexive polygons]{The $16$ reflexive
    polygons. $F_i$ and $F_{17-i}$ are dual for $i=0,\dots,6$, and
    self-dual for $i=7,\dots,10$. The corresponding toric surfaces are
    also known as $F_1=\CP^2$, $F_2=\CP^1\times\CP^1$, $F_3=dP_1$,
    $F_4=\CP^2[1,1,2]$, $F_5=dP_2$, $F_7=dP_3$, $F_{10}=\CP^2[1,2,3]$,
    where $dP_n$ are the del Pezzo surfaces obtained by blowing up
    $\mathbb{P}^2$ at $n$ points. Vertices defining toric sections are
    colored red. First derived in Figure~1 of \cite{2012arXiv1201.0930G}.}
  \label{fig:polygons}
\end{figure}

In this section, we contrast this with the Abelian gauge group factors
arising from F-theory compactifications on elliptically fibered
Calabi-Yau manifolds. For the purposes of this paper, we only consider
hypersurfaces in toric varieties such that the ambient toric variety
is itself fibered. In this case the toric ambient fiber is
two-dimensional and generically\footnote{Over the discriminant, the
  fiber will not be a toric variety but a reducible union of
  two-dimensional toric varieties.} again a toric variety. In terms of
toric data, the entire toric ambient space is always specified by a
reflexive polytope and the fibration amounts to a projection onto the
base fan. The preimage of the origin must be a reflexive polygon, and
this is the polygon determining the generic fiber of the fibration. In
\autoref{fig:polygons} we list the $16$ inequivalent reflexive
polygons, all of which can arise as the fiber polygon. If the
classification of $U(1)$ symmetries were to proceed analogously to
that of non-Abelian gauge symmetries, then one would expect the fiber
to fix the number of $U(1)$ factors uniquely. However, this is not the
case. The easiest way to illustrate this fact is to find concrete
counter-examples, and, indeed, these can easily be found. Using the
general methods discussed in later sections of this paper, we
constructed all different Calabi-Yau fourfolds with non-Abelian gauge
group $SU(5)$ and base manifold $B = \CP^3$.  In \autoref{t:n_U1s} we
list the different Abelian gauge groups we found for each of the $16$
fiber polygons. As they are not all the same, the total number of
$U(1)$ gauge factors is only captured by \emph{global} properties of
the Calabi-Yau geometry\footnote{Note that, apart from non-toric
  $U(1)$s, there can also be additional non-toric non-Abelian gauge
  groups. That is, gauge groups localized at discriminant components
  that are not defined by the vanishing of a single homogeneous
  coordinate. In particular, \emph{all} $SU(5)$ models with fiber
  $F_{14}$ and base $\mathbb{P}^3$ turn out to have such additional
  non-Abelian gauge factors.}.
\begin{table}
 \centering
 \begin{tabular}{c|cccccccccccccccc}
   Fiber polygon &
    $F_{1}$ &
    $F_{2}$ &
    $F_{3}$ &
    $F_{4}$ &
    $F_{5}$ &
    $F_{6}$ &
    $F_{7}$ &
    $F_{8}$ &
    $F_{9}$ &
    $F_{10}$ &
    $F_{11}$ &
    $F_{12}$ &
    $F_{14}$ 
    \\
    \hline
    $r_\text{min}$ &
    $0$ &
    $1$ &
    $1$ &
    $0$ &
    $2$ &
    $1$ &
    $3$ &
    $1$ &
    $2$ &
    $0$ &
    $1$ &
    $2$ &
    $1$
    \\
    $r_\text{max}$ &
    $4$ &
    $3$ &
    $4$ &
    $2$ &
    $4$ &
    $4$ &
    $3$ &
    $2$ &
    $3$ &
    $0$ &
    $1$ &
    $2$ &
    $1$  \\
  \end{tabular}
  \caption[Number of $U(1)$ factors for $SU(5)$ models over $\CP^3$]{
    Abelian $U(1)^r$ factors for all Calabi-Yau elliptic fibrations
    $X$ with $SU(5)$ 
    tops and fixed base manifold $B = \CP^3$. The 
    minimal number $r_\text{min} = \rank \MW_T(X)$ is the number of toric
    $U(1)$ factors. All values $r_\text{min}\leq r \leq r_\text{max}$ are
    realized by a fourfold.}
  \label{t:n_U1s}
\end{table}
Nevertheless, there are some properties that are fixed by the ambient
space fiber alone and we now focus on these.

A \emph{toric section} of an elliptically fibered Calabi-Yau
hypersurface is one that is obtained by setting one of the fiber
coordinates equal to zero. The conditions for a toric section
were first derived in~\cite{Rohsiepe:2005qg, 2012arXiv1201.0930G}, but for completeness
we restate them here. If we denote the fiber coordinates by $f_i$,
then the divisor $V(f_i)$ defines a section if it cuts out a single
point in the fiber, that is, the intersection product between $V(f_i)$
and the elliptic fiber $E$ must be
\begin{equation}
  [V(f_i)] \cap [E] = 1
  . 
  \label{e:section}
\end{equation}
The genus-one curve $E$ can be thought of as the one-dimensional
Calabi-Yau hypersurface in the fiber toric surface $F$, that is, the
anti-canonical hypersurface
\begin{equation}
  [E] = - c_1(F) = \sum_i [V(f_i)]. 
\end{equation}
We can always assume that the fiber fan is smooth. In this case, we
have $[V(f_i)] \cap [V(f_j)] = \delta_{i,j-1} + \delta_{i,j+1}$ for $i
\neq j$ and, therefore, eq.~\eqref{e:section} implies that a toric
section $V(f_i)$ must satisfy
\begin{equation}
  [V(f_i)] \cap [V(f_i)] = -1
  . \label{e:section2}
\end{equation}
To translate this into the geometry of the fan, let us denote the ray
corresponding to the toric coordinate $f_i$ by $v_i$. Then
\eqref{e:section2} is satisfied if the lattice spanned by the edges
connecting $v_i$ with its neighboring rays,
\begin{equation}
  N_i =
  \Span \left( v_i-v_{i-1},~ v_i-v_{i+1} \right)
\end{equation}
is the same as the fan lattice $N$, i.e.
\begin{equation}
  V(f_i) \text{ is a section } \Longleftrightarrow N_i = N.
\end{equation}
In particular, only vertices of a fiber polygon can give rise to toric
sections. Given this simple geometric prescription, one can easily
read off the toric sections of a given fiber polygon. In
\autoref{fig:polygons}, the vertices of the fiber polygon
corresponding to sections are shown in red.
\begin{table}[htbp]
  \vspace{-1cm}
  \renewcommand{\arraystretch}{0.7}
  \centering
  \begin{tabular}{c@{$\qquad$}r@{$\,\simeq\,$}l@{$\qquad$}c@{$\qquad$}c}
    Fiber polygon & \multicolumn{2}{c}{Toric sections} & Relations & $\MW_T$ \\
    \toprule
    $F_3$ & $(1,1) $&$ f_0$ & & $0$\\
    \midrule
    \multirow{3}{*}{$F_5$} & $(0, -1)$&$f_0$ & 
    & \multirow{3}{*}{$\mathbb{Z} \oplus \mathbb{Z}$}\\
    & $(-1,-1) $&$ f_1$ && \\
    & $(-1,0) $&$ f_2$ && \\
    \midrule
    \multirow{2}{*}{$F_6$} & $(0,1) $&$ f_0$ & & 
    \multirow{2}{*}{$\mathbb{Z}$}\\
    & $(-1,1) $&$ f_1$ & & \\
    \midrule
    \multirow{6}{*}{$F_7$} & $(-1,-1) $&$ f_0$ & 
    \multirow{3}{*}{$\sigma_1 = \sigma_3+\sigma_4$} &
    \multirow{6}{*}{$\mathbb{Z} \oplus \mathbb{Z} \oplus \mathbb{Z}$}\\
    & $(0,-1) $&$ f_1$ & & \\
    & $(1,0) $&$ f_2$ &  & \\
    & $(1,1) $&$ f_3$ & 
    \multirow{3}{*}{$\sigma_5 = \sigma_2 + \sigma_3$} & \\
    & $(0,1) $&$ f_4$ &  & \\
    & $(-1,0) $&$ f_5$ & & \\
    \midrule
    \multirow{2}{*}{$F_8$} & $(-1,1) $&$ f_0$ &  & 
    \multirow{2}{*}{$\mathbb{Z}$}\\
    & $(1,1) $&$ f_1$ &&\\
    \midrule
    \multirow{4}{*}{$F_9$} & $(-1,-1) $&$ f_0$ & 
    \multirow{4}{*}{$\sigma_1 = \sigma_2+\sigma_3$} 
    & \multirow{4}{*}{$\mathbb{Z} \oplus \mathbb{Z}$}\\
    & $(0,-1) $&$ f_1$ &&\\
    & $(0,1) $&$ f_2$ & & \\
    & $(1,1) $&$ f_3$ & & \\
    \midrule
    $F_{10}$ & $(-3,-2) $&$ f_0$ &  &  $0$ \\
    \midrule
    \multirow{3}{*}{$F_{11}$} & $(-1,-1) $&$ f_0$ & & 
    \multirow{3}{*}{$\mathbb{Z}$}\\
    & $(0,-1) $&$ f_1$ &  $\sigma_1 = 2 \sigma_2$ & \\
    & $(-1,2) $&$ f_2$ & & \\
    \midrule
    \multirow{5}{*}{$F_{12}$} & $(-1,-1) $&$ f_0$ &  & 
    \multirow{5}{*}{$\mathbb{Z} \oplus \mathbb{Z}$}\\
    & $(1,-1) $&$ f_1$ & $\sigma_1 = \sigma_3 + \sigma_4$ &\\
    & $(1,0) $&$ f_2$ &  & \\
    & $(0,1) $&$ f_3$ & $\sigma_4 = \sigma_1 + \sigma_2$ & \\
    & $(-1,1) $&$ f_4$ & & \\
    \midrule
    \multirow{2}{*}{$F_{13}$} & 
    $(-1,-2) $&$ f_0$ & 
    \multirow{2}{*}{$2 \sigma_1 = 0$} &
    \multirow{2}{*}{$\mathbb{Z}_2$}\\
    & $(-1,2) $&$ f_1$ &  &\\
    \midrule
    \multirow{4}{*}{$F_{14}$} & 
    $(-1,0) $&$ f_0$ & 
    \multirow{2}{*}{$\sigma_1 = 2 \sigma_2$} & 
    \multirow{4}{*}{$\mathbb{Z}$}\\
    & $(-1,2) $&$ f_1$ &  & \\
    & $(2,-1) $&$ f_2$ & 
    \multirow{2}{*}{$\sigma_3 = \sigma_1 + \sigma_2$} & \\
    & $(0,-1) $&$ f_3$ & & \\
    \midrule
    \multirow{4}{*}{$F_{15}$} & 
    $(-1,-1) $&$ f_0$ & 
    \multirow{2}{*}{$2 \sigma_2 = 0$} & 
    \multirow{4}{*}{$\mathbb{Z} \oplus \mathbb{Z}_2$}\\
    & $(1,-1) $&$ f_1$ &  &\\
    & $(1,1) $&$ f_2$ & \multirow{2}{*}{$\sigma_3 = \sigma_1 + \sigma_2$} & \\
    & $(-1,1) $&$ f_3$ & & \\
    \midrule
    \multirow{3}{*}{$F_{16}$} & 
    $(-1,-1) $&$ f_0$ & 
    $ 3 \sigma_1 = 0$ & 
    \multirow{3}{*}{$\mathbb{Z}_3$}\\
    & $(2,-1) $&$ f_1$ & & \\
    & $(-1,2) $&$ f_2$ & $ \sigma_2 = 2 \sigma_1$ & \\
    \bottomrule
  \end{tabular}
  \caption[Toric Mordell-Weil groups]{Toric sections corresponding to
    the reflexive polytopes and the 
    toric subgroup of the Mordell-Weil group generated by them.}
  \label{t:toric_sections}
\end{table}

Having defined toric sections, we now define the \emph{toric}
Mordell-Weil group $\MW_T$ if there is at least one toric
section. First, recall that the sections of an elliptic fibration form
an Abelian group using the group law on each elliptic curve
fiber. This group is the Mordell-Weil group of the compactification
manifold and plays an important role in F-theory
compactifications~\cite{Morrison:1996pp, Park:2011ji,
  Morrison:2012ei}. The rank of the Mordell-Weil group, measuring the
free part, equals the number of independent $U(1)$ factors. The
torsion subgroup are discrete symmetries in F-theory that might be
important for proton stability. Now, the toric sections form a finite
subset of all sections, but that subset is usually not closed under
the group law. However, if there is at least one toric section to use
as the zero-section, then they generate a subgroup which we call the
\emph{toric} Mordell-Weil group
\begin{equation}
 \MW_T = \langle f_i-f_j \rangle  \subset \MW.
\end{equation}
Hence, once the toric sections of a certain fiber polygon have been
found one has to determine the relations among them.  Clearly, the
relations only need to be computed at a single sufficiently generic
fiber. As we have just reviewed, a toric section $f_i =0$ cuts out
precisely one point on a generic elliptic curve contained in the fiber
variety. Moreover, this point is rational if we choose the
hypersurface to have rational coefficients and a rational point on the
base. Note that an elliptic fibration is just an elliptic curve over
the function field in the base, so the notions of sections and group
law equally apply to the set of rational points $E(\Q)$ on an elliptic
curve $E$. The analogue of a toric section is a toric rational point,
that is, a point where one of the homogeneous variables is equal to
zero. We review the geometric group law in \autoref{a:curves}. The
generic Mordell-Weil group $E(\Q)$ obtained in this way only depends
on the fiber polygon, but not on the details of the embedding in a
higher-dimensional reflexive polytope. However, note that special
fibers, that is, over special points in the base or the restriction of
non-generic hypersurface equations will in general have a different
$E(\Q)$.

We now apply this to the $16$ toric surfaces that can arise as the
ambient space of the generic fiber. We first pick a toric section,
say $f_0$ in the notation of \autoref{t:toric_sections}, as the zero
section. The other sections, let us call them
\begin{equation}
  \sigma_i = f_i - f_0, \qquad i > 0 ,
\end{equation}
then satisfy relations among them which determine the toric
Mordell-Weil group $\MW_T(X) \subseteq \MW(X)$. All fiber polygons
with sections contain the $\CP^2$ polygon $F_1$, so all of their
Calabi-Yau hypersurface equations are specializations of the cubic in
$\CP^2$. By picking a sufficiently generic fiber $E$ and applying the
$9:1$ map from the cubic to the Weierstrass equation, we map the $f_i$
to specific rational points on the Weierstrass model of $E$. There,
the group structure on $E(\Q)$ can easily be determined. Comparing the
generators for $E(\Q)$ with the restrictions $\sigma_i|_E$, we obtain
relations between the $\sigma_i|_E$. Up to $3$-torsion, these must
also hold in the toric Mordell-Weil group $\MW_T(X)$ of the Calabi-Yau
manifold $X$. However, there can be extra factors of $\Z_3$ owing to
the fact that the map to the Weierstrass form includes a $3^2:1$
isogeny.\footnote{For example, the three toric sections in $F_{16}$
  all map to the same section of the Weierstrass model.} Using the
geometric group law in the actual fiber $E$, we have verified which
relations hold directly and which hold only modulo three-torsion. The
result are the actual relations between the toric sections listed in
\autoref{t:toric_sections}. Incidentally, and somewhat disappointingly
for phenomenology purposes, the three polygons whose toric
Mordell-Weil groups contain non-trivial torsion subgroups turn out to
be precisely those that do not support an $SU(5)$ top.

Finally, let us stress that from a physical point of view, the
presence of non-toric sections poses no problems. In practice, a vast
number of torically constructed Calabi-Yau manifolds has such sections
and while their defining equations may not be as easily found as in
the toric case, for practical purposes it often suffices to know their
homology classes~\cite{Braun:2013yti}.


%% file: AllTops.tex
\section{All \boldmath$SU(5)$ Tops}
\label{sec:tops}

By definition, a \emph{top}~\cite{Candelas:1996su, Candelas:2012uu} is
the preimage of a base ray in a toric morphism where the Calabi-Yau
hypersurface (in the total space) is a genus-1 fibration. This
preimage is a lattice polytope with one facet, namely the kernel of
the projection, passing through the origin. This special facet passing
through the origin is the fiber polygon. It is again reflexive if the
total space polytope is, and hence must be one of the $16$ polygons
shown in \autoref{fig:polygons}. For example, consider an elliptically
fibered K3 constructed as a hypersurface in a three-dimensional
fibered toric variety. Then the three-dimensional reflexive polytope
is split into two tops by the projection to the $\CP^1$-fan. The top
determines the structure of the toric variety over the base divisor
specified by the base ray. In particular, the top often enforces a
particular Kodaira fiber in the hypersurface.

In the following, we will be particularly interested in elliptic
fibrations with $I_5$ Kodaira fibers, leading to $SU(5)$ GUTs. In
general, the $SU(n)$ tops are spanned by two parallel facets lying in
two lattice planes of distance one. For convenience, let us pick
coordinates $(x,y,z)$ such that the fiber polygon vertices are at
$z=0$ and all remaining vertices at height $z=1$. The
circumference\footnote{That is, the number of lattice points on the
  boundary.} of the facet at $z=1$ equals $n$. Hence, an $SU(5)$ top
is defined by the fiber polygon together with a polygon with
circumference $5$~\cite{Perevalov:1997vw, Candelas:1997pq}. The
defining data for all $37$ $SU(5)$-tops is shown in
\autoref{fig:tops}.\footnote{Note that symmetries of the tops are
  divided out. For example, the $F_5$ fiber polygon is symmetric under
  a $\Z_2$ automorphism. It supports $5$ distinct tops, of which one
  is symmetric under the $\Z_2$ and four are not. If one were to
  disregard the symmetry, one would arrive at the $9$ tops
  of~\cite{Mayrhofer:2012zy}.} Note that the $GL(3,\Z)$-subgroup
generated by $(x,y,z)\mapsto (x+\alpha z, y+\beta z, z)$ still acts on
the tops after fixing the fiber polygon and therefore the $x$, $y$
coordinates of the facet at $z=1$ can be shifted arbitrarily. This is
why in the origin is only marked for the fiber polygons in
\autoref{fig:tops}, but not for the polygons at height one.

There is another useful notation for tops. The dual polyhedron
$\tau^*$ of a top $\tau$ is a non-compact semi-infinite prism whose
cross-section is the dual fiber polygon $F^*$. One of the dual vertices
is $(x^*, y^*, z^*)=(0,0,-1)$, the dual of the facet at $z=1$. The
$z^*$ value at which the semi-infinite prism is cut off defines a
function on the remaining lattice points $F^*\cap \Z^2$ of the dual
fiber polygon. This function $z^*: \partial F^*\cap \Z^2 \to \Q$
defines the top as
\begin{equation}
  \tau^* = \conv
  \Big\{ \big(p_x, p_y, z^*(p)\big) 
  ~\Big|~ 
  p \in \partial F^*\cap \Z^2 \Big\}
  + 
  \R_{\geq} \cdot (0,0,1)
  ,
\end{equation}
which is an equivalent notation to specify a top. Again, there is a
residual $GL(3,\Z)$ symmetry action on the coordinate choices, since
two functions $z^*$ that differ only by a linear function define the
same top. Shifting by a linear function, one can always bring $z^*$ in
a form where $z^*(p)\geq -1$ for all boundary points and this
convention has been used in \autoref{fig:tops}.
\begin{figure}[p]
  \vspace{-1cm}
  \centering
  \begin{tabular}{m{3cm}|m{12cm}}
    \includegraphics{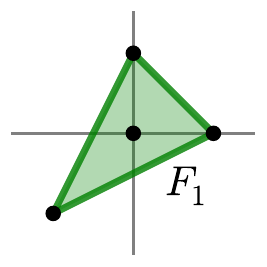} &
    \includegraphics{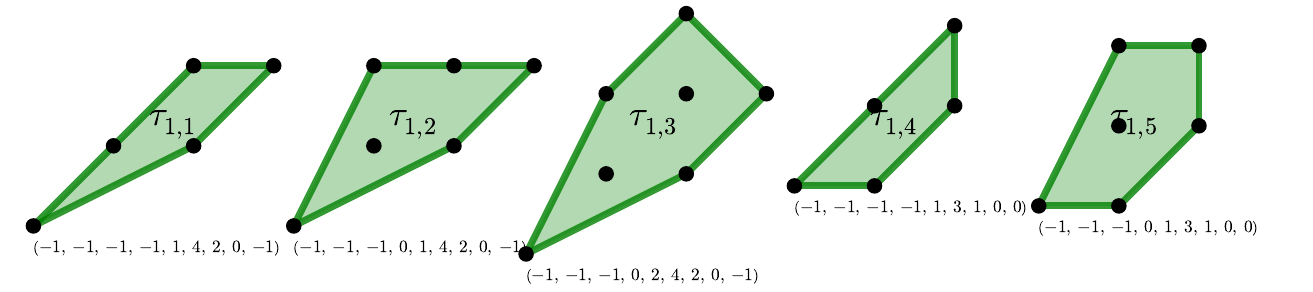}
    \\\hline
    \includegraphics{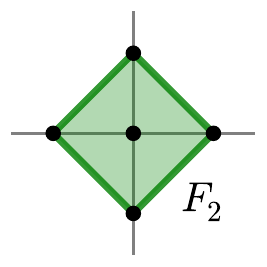} &
    \includegraphics{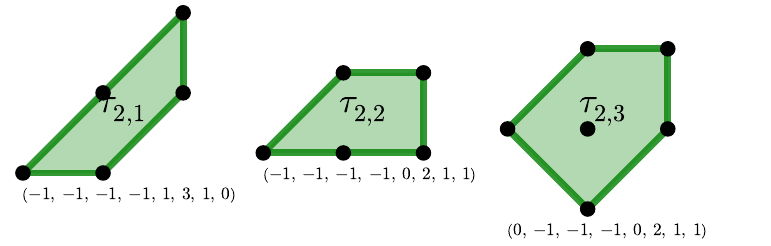}
    \\\hline
    \includegraphics{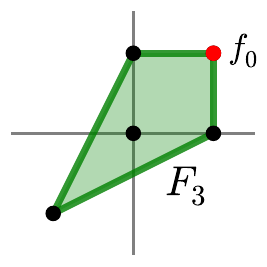} &
    \includegraphics{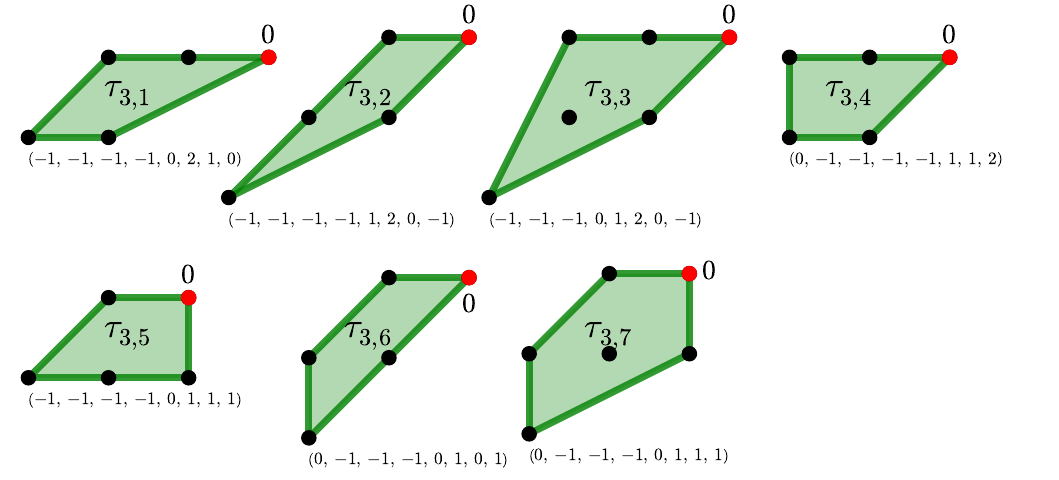}
    \\\hline
    \includegraphics{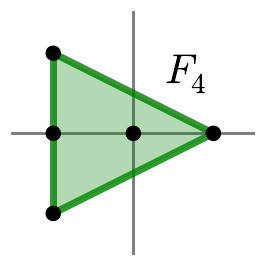} &
    \includegraphics{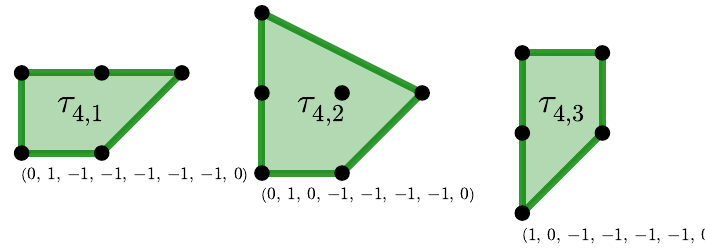}
    \\\hline
    \includegraphics{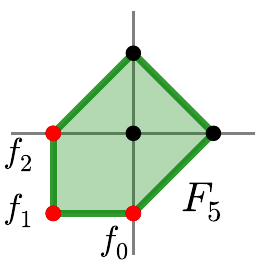} &
    \includegraphics{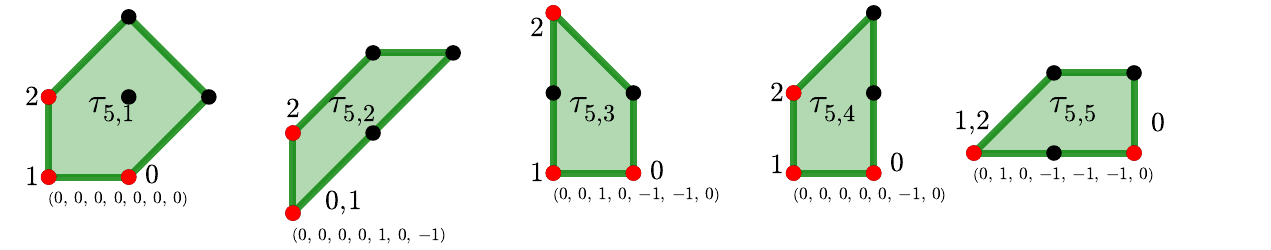}
    \\\hline
    \includegraphics{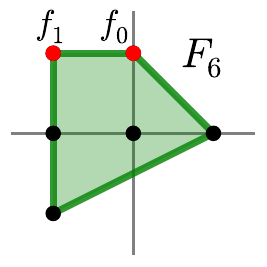} &
    \includegraphics{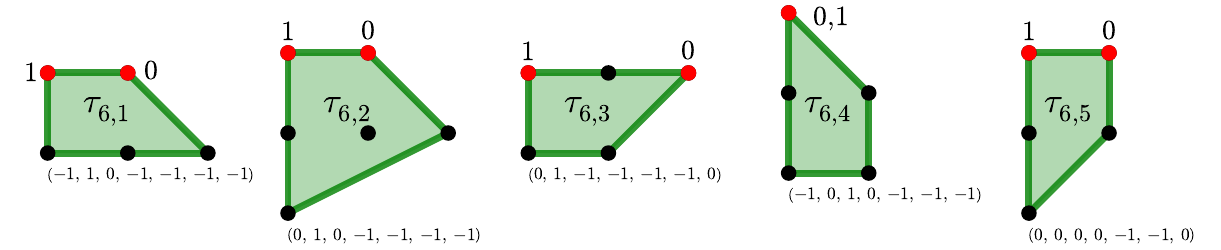}
  \end{tabular}
  \caption[All $SU(5)$ tops]{The $SU(5)$ tops based on the $16$
    reflexive polygons. Numbers next to boundary points of the facet
    in the $z=1$ plane indicate which toric sections intersect the
    associated exceptional divisor.}
  \label{fig:tops}
\end{figure}
\captionsetup[figure]{list=no}
\begin{figure}[p]
  \ContinuedFloat
  \vspace{-1cm}
  \hspace{-1cm}
  \begin{sideways}
    \centering
    \begin{tabular}{m{4cm}|m{4cm}}
    \includegraphics{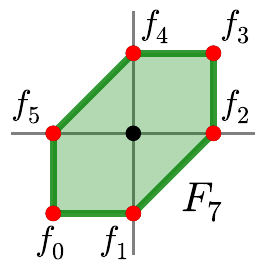} &
    \includegraphics{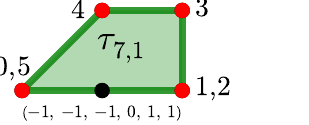}
    \\\hline
    \includegraphics{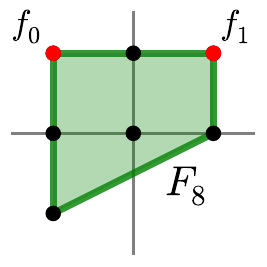} &
    \includegraphics{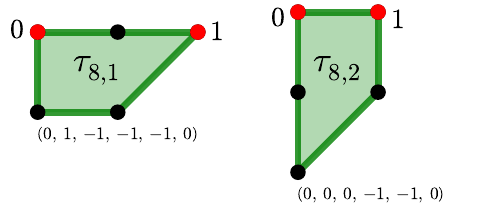}
    \\\hline
    \includegraphics{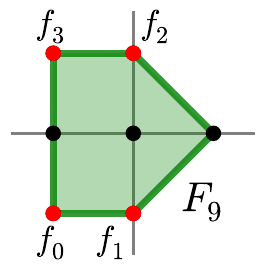} &
    \includegraphics{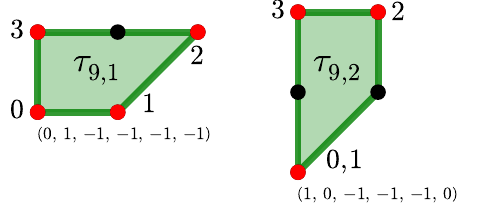}
    \\\hline
    \includegraphics{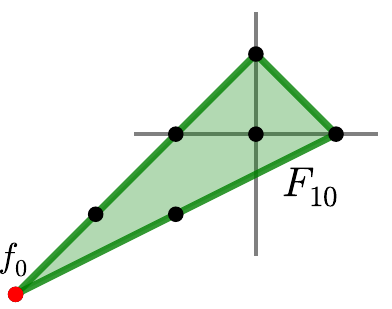} &
    \includegraphics{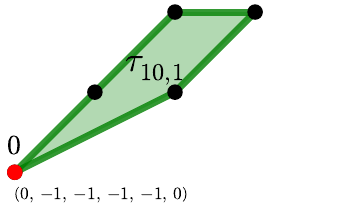}
    \\\hline
    \includegraphics{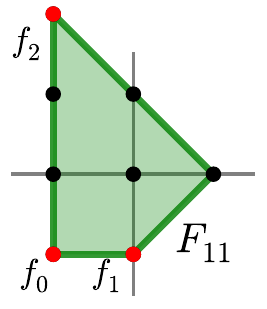} &
    \includegraphics{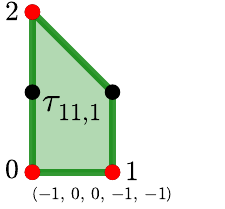}
  \end{tabular}
  \hspace{2cm}
    \begin{tabular}{m{3cm}|m{2cm}}
    \includegraphics{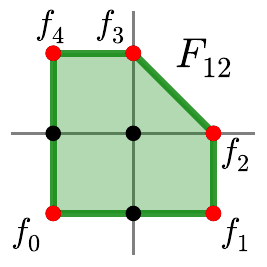} &
    \includegraphics{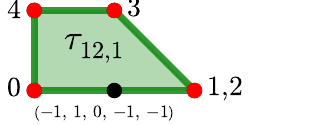}
    \\\hline
    \includegraphics{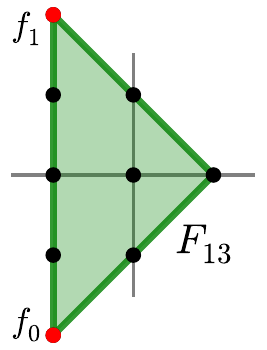} &
    \\\hline
    \includegraphics{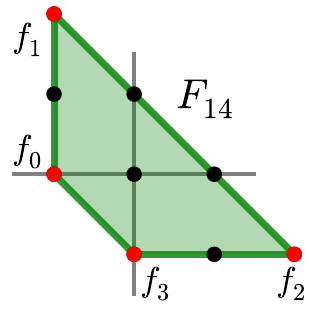} &
    \includegraphics{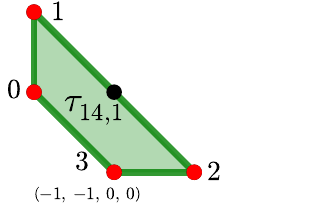}
    \\\hline
    \includegraphics{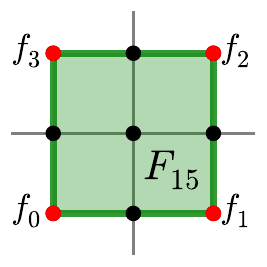} &
    \\\hline
    \includegraphics{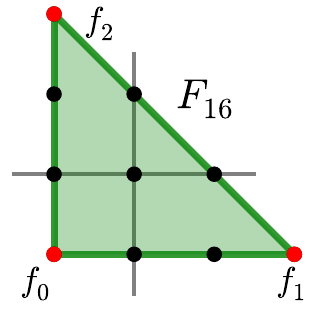} &
  \end{tabular}
  \end{sideways}
  \caption{(continued) The $SU(5)$ tops $\tau_{i,j}$ based on the $16$
    reflexive polygons. For each reflexive polygon (the fiber polygon
    at $z=0$), the admissible facets at $z=1$ are listed. Below each
    the values of $z^*$ on the vertices of the dual polygon (in
    clockwise order, starting at the ``y''-axis) are given, which also
    an equivalent way of specifying the top. See discussion at the
    beginning of \autoref{sec:tops}.}
  \label{fig:tops_continued}
\end{figure}
\captionsetup[figure]{list=yes}

In general, one can use a top to construct a local Kodaira fiber. This
is the most general definition of a top, and has been used
in~\cite{Bouchard:2003bu}. Of course, not every local Kodaira fiber can
appear in a compact elliptic fibration. However, all of the $SU(5)$
tops actually appear in elliptic K3 fibrations, that is, the tops
occur in the list of $1052$ half-K3 polytopes~\cite{Candelas:1996su}.


%% file: U1charges.tex
\section{\boldmath Determining $U(1)$ Charges} 
\label{sec:U(1)charges}

Having identified all \emph{toric} Mordell-Weil generators for the
$16$ two-dimensional reflexive polytopes, an obvious question is
whether one can make any general statements about matter
representations charged under the $U(1)$ gauge groups corresponding to
the Mordell-Weil generators. Similar to \autoref{sec:fiber}, it turns
out that information about toric $U(1)$s is already contained in the
top alone~\cite{Braun:2013yti, Borchmann:2013jwa}.  For illustration,
let us restrict ourselves to gauge groups such that the non-Abelian
factor is $SU(5)$.  However, nothing really depends on this choice and
our reasoning can readily be extended to other gauge groups. Generic
F-theory compactifications only give rise to two distinct non-trivial
$SU(5)$ representations, $\Rep{5}$ and $\Rep{10}$. In particular, this
is the case for generic hypersurfaces in toric varieties. This is
because, generically, the rank $4$ gauge group will only be enhanced
to rank $5$, that is, $SU(6)$ and $SO(10)$. The matter fields then
arise from decomposing their adjoint representations. We begin by
studying the constraints on the $U(1)$ charges of the $\Rep{5}$
representation before proceeding with the latter representation.

\subsection{Different Splits and the Fundamental Representation}

Let us now examine the possible $U(1)$ charges of matter in the
$\Rep{5}$ representation in a systematic manner. In many ways, this is
a straightforward generalization of the example analyzed
in~\cite{Braun:2013yti} for the case of a single $U(1)$ and we
therefore reiterate the arguments made in the above paper in the next
paragraphs.

By definition, a section cuts out a single point in the fiber of our
elliptically fibered Calabi-Yau manifold over a dense subset of the
base manifold $B$.\footnote{If one requires the section to
  be holomorphic, then this is in fact true for all of $B$.
  Rational sections, however, may wrap entire fiber components over
  certain lower-dimensional loci of the base,
  see~\cite{Morrison:2012ei, Mayrhofer:2012zy, Braun:2013yti,
    Grimm:2013oga} for related discussions.} Over generic points of
the GUT divisor, however, the torus fiber degenerates. In the case of
an $SU(5)$ gauge group, it splits into an $I_5$ Kodaira fiber
consisting of $5$ $\CP^1$ components intersecting as the affine Dynkin
diagram $\tilde{A}_4$. There is no reason for the sections of the
compactification manifold to mark points on the same $\CP^1$, and, in
general, they will intersect different $\CP^1$ irreducible
components. One is always free to designate any one of the sections to
be the \emph{zero section}, that is, one takes the point marked by this
section to be the identity element on the elliptic curve. Physically,
the closed two-form corresponding to the zero section gives rise to
the Kaluza-Klein vector $A^0_{\mu}$ in the expansion of the RR
three-form. By choosing a zero section, one implicitly fixes the
affine node of $\tilde{A}_4$ as the irreducible component of the $I_5$
Kodaira fiber that the zero section intersects. 

\begin{figure}[htb]
  \centering
  \includegraphics[width=0.8\textwidth]{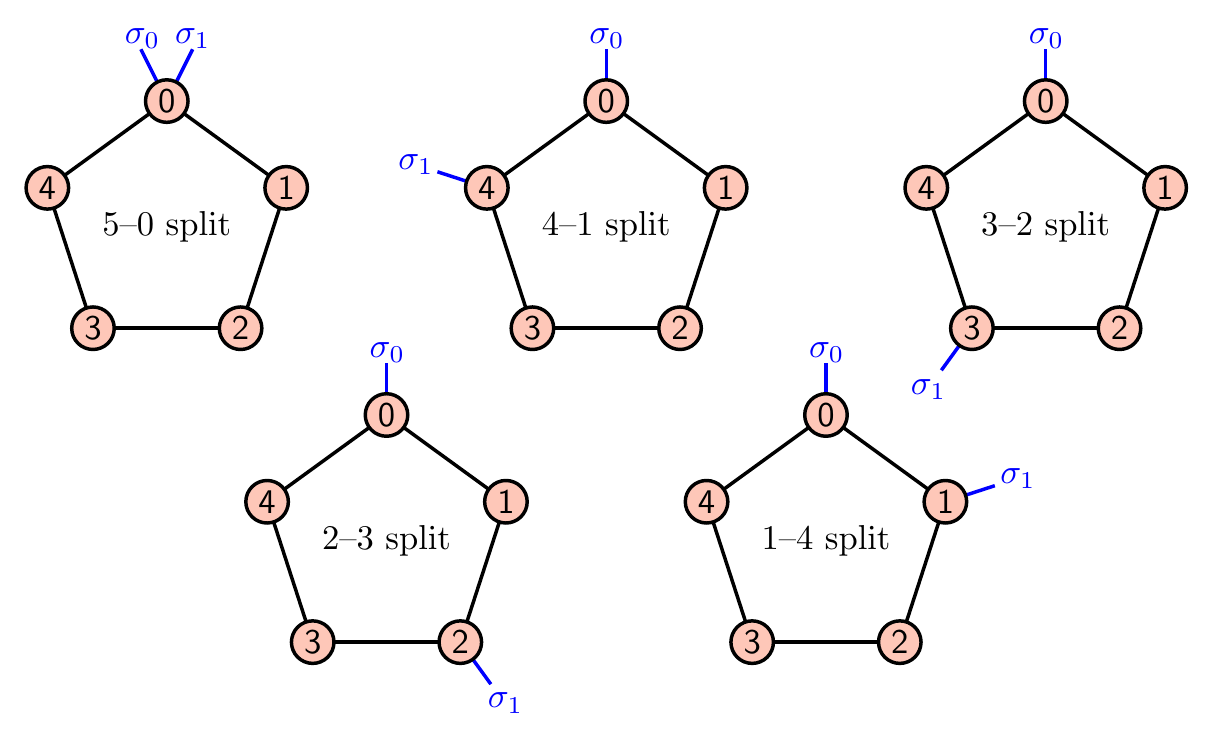}
  \caption[The different splits of the $SU(5)$]{The different splits
    for the case in which $\sigma_0$ denotes the zero section and
    $\sigma_1$ is one of possibly more independent Mordell-Weil
    generators. In the case of a single $U(1)$, the $i$-$(5-i)$-split
    and the $(5-i)$-$i$-split are equivalent under to the $\Z_2$ outer
    automorphism of $\mathfrak{su}(5)$.}
  \label{fig:splits}
\end{figure}

Having designated a zero section, one then identifies the remaining
$\CP^1$ components with the simple roots of $\mathfrak{su}(5)$ such
that their intersections reproduce the inner products of the
associated coroots. Note that identifying the $\CP^1$ components with
simple roots of $\mathfrak{su}(5)$ is unique only up to a $\Z_2$
ambiguity corresponding to the outer automorphism $D_i \leftrightarrow
D_{5-i}$. After eliminating said ambiguity, one is left with only
three possible intersections patterns in the case of a single $U(1)$
generator, namely the first three of \autoref{fig:splits}.
The physics behind these three different intersection patters is that
they constrain the charges of the $\Rep{5}$ representation up to
multiples\footnote{More precisely, the charges are determined up to a
  multiple of $5$ after rescaling the naive $U(1)$ generator, obtained
  by applying the Shioda map~\cite{shioda1989mordell, 
    shioda1990mordell} to the Mordell-Weil generator,
  by a factor of $5$ to make all charges integral.} of $5$ in the
non-trivial cases, that is, except for the $0$-$5$ split. Concretely,
the relation is
\begin{equation}
  U(1) 
  \text{-generator with $i$-$(5-i)$-split}
  \quad \Leftrightarrow \quad 
  Q_{U(1)}(\Rep{5}) \equiv i \tmod 5
  .
\end{equation}
This interpretation of the intersection pattern generalizes easily to
the case of multiple Abelian gauge group factors.  Having fixed a zero
section and one of the two choices for assigning simple roots of
$\mathfrak{su}(5)$ to the irreducible fiber components, one can
compute the intersection numbers between the fiber $\CP^1$s and any
given Mordell-Weil generator $\sigma_m$. For sections $\sigma_j$ with
$j = 1, \dots,n_{U(1)}$ intersecting the $i_j$-th fiber component we
then find that all our matter representations have $U(1)$ charges
satisfying
\begin{equation}
 Q_{U(1)_j}(\Rep{5}) \equiv i_j \tmod 5
 .
\end{equation}
The intersection structure between toric sections and the irreducible
fiber components is already fixed by the top alone. In fact, one can
easily read off the intersection numbers from the geometry of the
top. As reviewed in \autoref{sec:tops}, a top is a three-dimensional
polyhedron such that the origin is interior to a facet. In the case of
$SU(n)$ gauge groups, it consists of two parallel polygons. One of
these, the fiber polygon, is the facet containing the origin. It is a
reflexive sub-polytope. The other polygon lies at height $1$ and its
integral boundary points reproduce the affine Dynkin diagram
$\tilde{A}_n$. Now, let $v$ be a vertex of the fiber polygon
corresponding to a toric section $\sigma$. Furthermore, let us denote
the lattice points corresponding to the $i$-th fiber component by
$w_i$. Then $\sigma$ intersects the $i$-th fiber component if and only
if $v$ and $w_i$ share an edge.
\begin{figure}[htb]
  \centering
  \includegraphics[width=0.9\textwidth]{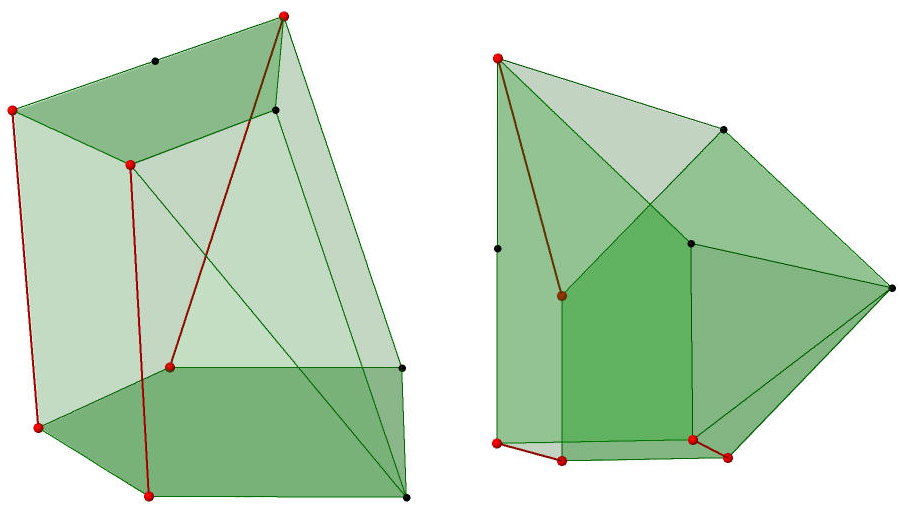}
  \caption[The $\tau_{5,3}$ top]{Two different 3D visualizations of the
    entire top $\tau_{5,3}$. Fiber vertices corresponding to sections
    and lattice points associated to exceptional divisors intersecting
    them are colored red, as are the edges connecting them.}
  \label{fig:top3d}
\end{figure}
Using this prescription, we have determined the intersection numbers
for all $SU(5)$ tops in \autoref{fig:tops} by listing the sections
intersecting a certain exceptional divisor next to the corresponding
lattice point of the $z=1$ facet of the top. To give an example,
consider the top $\tau_{5,3}$. From \autoref{t:toric_sections}, we see
that the toric sections generate a subgroup $\MW_T \cong \Z \oplus \Z
\subseteq \MW$ of the entire Mordell-Weil group. Now, pick $f_0$ as
the zero section and assign simple roots $\alpha_i$ in clockwise order
to the boundary points of $\tau_{5,3}$. Taking $f_1$ and $f_2$ as
sections generating $U(1)_1$ and $U(1)_2$, we find that the charges of
the $\Rep{5}$ representations must satisfy
\begin{equation}
 Q_{U(1)_1}(\Rep{5}) \equiv 1 \tmod 5 
 \quad \text{and} \quad 
 Q_{U(1)_2}(\Rep{5}) \equiv 3 \tmod 5
 .
 \label{e:conditions_5s_example}
\end{equation}
\begin{table}
  \centering
  \begin{tabular}{|c||cc|ccc|}
    \hline
    $\nabla_{5,3}(-3, -3)$ & 
    \multicolumn{2}{c|}{fiber} & 
    \multicolumn{3}{c|}{base}
    \\ \hline \hline
    \multirow{4}{*}{fiber}
    &  1 &  0 & 0 & 0 & 0 \\ 
    &  0 &  1 & 0 & 0 & 0 \\
    & -1 &  0 & 0 & 0 & 0 \\ 
    &  0 & -1 & 0 & 0 & 0 \\ 
    & -1 & -1 & 0 & 0 & 0 \\  \hline
    \multirow{5}{*}{$\tau_{5,3}$}
    &  0 & -1 & 1 & 0 & 0 \\ 
    & -1 & -1 & 1 & 0 & 0 \\
    & -1 &  0 & 1 & 0 & 0 \\ 
    & -1 &  1 & 1 & 0 & 0 \\
    &  0 &  0 & 1 & 0 & 0 \\  \hline
    \multirow{1}{*}{trivial top}
    &  0 &  0 & 0 & 1 & 0 \\ \hline 
    \multirow{1}{*}{trivial top}
    &  0 &  0 & 0 & 0 & 1 \\  \hline
    \multirow{1}{*}{trivial top}
    &$-3$&$-3$&-1 &-1 &-1 \\  \hline
  \end{tabular}
  \caption[All completions of the $\tau_{5,3}$ top to a Calabi-Yau
  fourfold over $\CP^3$]{Toric data of one of $30$ inequivalent completions of 
    $\tau_{5,3}$ to the polytope of a Calabi-Yau
    fourfold with base $\CP^3$.}
  \label{t:example_5s}
\end{table}
Using the algorithm described in the next section, one can complete
this top to a five-dimensional reflexive polyhedron. For simplicity,
we choose the base manifold to be $\CP^3$ and find that there are $30$
inequivalent embeddings, one of which we list in
\autoref{t:example_5s}. Independent of the chosen triangulation, one
finds that a generic hypersurface inside this space possesses three
distinct curves in the base over which the $SU(5)$ singularity is
enhanced to $SU(6)$, giving rise to matter in the $\Rep{5}$
representation of $SU(5)$. Going through similar calculations as for
example in~\cite{Mayrhofer:2012zy, Braun:2013yti, Cvetic:2013nia}, one
finds that the $U(1)$ charges of these three curves with respect to
the generators associated with $(-1,-1,\mathbf{0})$ and
$(-1,0,\mathbf{0})$ are
\begin{equation}
  \Rep{5}_{-4,-2}
  ,~
  \Rep{5}_{1,-2}
  ,~\text{and}~
  \Rep{5}_{1,3}
  ,
\end{equation}
in agreement with eq.~\eqref{e:conditions_5s_example}.

Finally, let us point out that the above notion of splits agrees with the cases 
that have been analyzed with the split spectral cover
constructions~\cite{Marsano:2009gv,Marsano:2009wr,Dudas:2010zb,Dolan:2011iu}
only in the case of a single $U(1)$. As soon as there are multiple
Abelian gauge symmetries, our notation describes the ``split'' between
the section generating the particular $U(1)$ symmetry and the zero section,
whereas the split spectral cover constructions denote by split
the factorization pattern of the spectral cover. Hence, when
there are multiple $U(1)$s we determine a split with respect
to each one of them.

\subsection{A No-Go Theorem for the Antisymmetric Fields}

We now turn to the $\Rep{10}$ matter fields of the $SU(5)$
GUT. Somewhat surprisingly, their geometric origin is rather different
from the $\Rep{5}$ matter fields. Recall that the $\Rep{5}$ matter
fields come from individual $\CP^1$ in the $I_5$ Kodaira fiber
degenerating into two irreducible components. This kind of
degeneration will never yield a codimension-two $I_1^*$ Kodaira fiber
where the $\Rep{10}$ matter field is localized: Splitting nodes of the
$I_5$ Kodaira fiber will never eliminate the fundamental group
$\pi_1(I_5)=\Z$ of the Kodaira fiber, but $\pi_1(I_1^*)=0$. The only
way to obtain a simply connected fiber is to have the hypersurface
equation vanish identically on a toric curve of the top. That is,
along the intersection of the irreducible components of the toric
surfaces in the fiber of the ambient toric variety. Note that the
irreducible components of the two-dimensional ambient space fiber
correspond to the vertices of the top that are not interior to a facet
and not part of the fiber polygon. They intersect in a toric curve
$\simeq \CP^1$ whenever the triangulation induced by the fan joins two
vertices.

\begin{figure}[htbp]
  \centering
  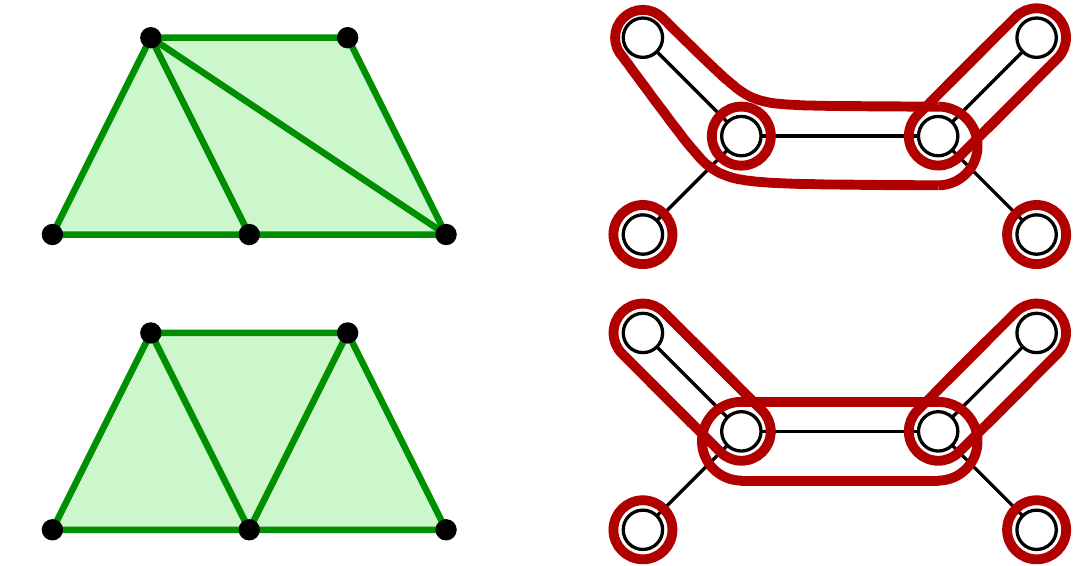
  \caption[Triangulations and $I_5 \rightarrow I_1^*$
  degenerations]{Left: The two possible fine triangulations of the
    lattice polygon at height $z=1$ in the $SU(5)$-top. Right: The
    corresponding degeneration of the $I_5 \rightarrow I_1^*$ Kodaira
    fiber.}
\label{fig:triang}
\end{figure}

As we will see in \autoref{sec:flat2}, if a $SU(5)$-top contains a
point interior to a facet then the fibration is not flat. This means
that the low-energy theory is not an ordinary gauge theory and we only
have to focus on tops without facet interior points. For an $SU(5)$ top
this means that the facet at height $z=1$ is a degenerate lattice
pentagon with one of the lattice points at a midpoint of an edge. Up
to isomorphism, there is only a single such lattice pentagon, see
\autoref{fig:tops}. There are two fine triangulations $T_1$ and $T_2$
of this boundary facet and they are shown on the left hand side of
\autoref{fig:triang}.
Regardless of the triangulation, the degenerate toric ambient space
fiber consists of $5$ irreducible surfaces $V(d_1)$, $\dots$,
$V(d_5)$. These always intersect cyclically in toric curves, that is,
$V(d_i)\cap V(d_{i+1}) \simeq \CP^1$. Depending on the triangulation,
they additionally intersect as the internal 1-simplices in the
triangulation, that is,
\begin{itemize}
\item Triangulation $T_1$: $V(d_1)\cap V(d_4) \simeq \CP^1$ and
  $V(d_1)\cap V(d_3)\simeq \CP^1$,
\item Triangulation $T_2$: $V(d_1)\cap V(d_4) \simeq \CP^1$ and
  $V(d_2)\cap V(d_4)\simeq \CP^1$.
\end{itemize}
The Calabi-Yau hypersurface generically intersects the toric $\CP^1$s
corresponding to the boundary 1-simplices in a point, and is a
non-zero constant on the toric $\CP^1$ corresponding to the internal
1-simplices. As argued in the beginning of this section, the
$\Rep{10}$ matter is localized when the whole toric $\CP^1$ is
contained in the hypersurface, that is, where the above constant
happens to be zero.\footnote{This is at a codimension-one curve of the
  discriminant, that is, it is of codimension two in the base.} Since the
internal $1$-simplices are internal to the same facet of the top,
the hypersurface always vanishes simultaneously on both toric curves.
These two toric $\CP^1$s intersect in a toric point, the
containing $2$-simplex. Hence they form two nodes joined by an edge
in the dual fiber diagram, which will turn out to be the middle two
nodes of the $\tilde{D}_5$ extended Dynkin diagram.

Intersecting the hypersurface $Y$ with the ambient space irreducible
surface components of a fiber yields additional curve components for
the degenerate elliptic fiber. These necessarily contain the toric
curves of the adjacent internal $1$-simplices as irreducible
components. For example, in triangulation $T_1$ the intersection
$Y\cap V(d_1)$ contains both toric surfaces $V(d_1)\cap V(d_4)$ and
$V(d_1)\cap V(d_3)$ as irreducible components. Likewise, $Y\cap
V(d_5)$ contains none of the toric $\CP^1$ since the vertex $d_5$ is
not adjacent to an interior $1$-simplex. This fixes the degeneration
of the $I_5$ Kodaira fiber, that is the $5$ curves $Y\cap V(d_i)$ away
from the matter curve, to be the one shown on the right hand side of
\autoref{fig:triang}.

This is the key observation: The triangulation of the top fixes the
degeneration of the codimension-one Kodaira fiber at the
codimension-two $\Rep{10}$ matter curves of a toric
hypersurface. Since the triangulation is fixed for a given manifold,
the degeneration is the same for all $\Rep{10}$. Importantly,
this behavior is different from that of the $\Rep{5}$ matter curves,
where different degenerations occur over different codimension-two fibers.
As a corollary, the $U(1)$ charges
of all $\Rep{10}$ matter representations are equal. In other words, if
one wants to construct F-theory GUTs such that the $\Rep{10}$ fields
carry different $U(1)$ charges then one needs to consider complete
intersections such that the fiber is at least codimension-two in the
ambient space fiber~\cite{Esole:2011cn,workinprogress}. Although we will not attempt
such a construction in this paper, let us note that the toric machinery applies
equally. The only difference is that one would construct a fibration
such that the toric ambient space fiber is higher dimensional, for
example, a reflexive 3-dimensional polytope instead of a reflexive polygon. Such
a complete intersection top is then a 4-dimensional polytope with one
facet being the 3-dimensional fiber polytope. The complete
intersection tops can be combined into reflexive polytopes as
described in \autoref{sec:compactification}.


%% file: fig_triangulation.pdf_t
\begin{picture}(0,0)%
\includegraphics{fig_triangulation.pdf}%
\end{picture}%
\setlength{\unitlength}{4144sp}%
\begingroup\makeatletter\ifx\SetFigFont\undefined%
\gdef\SetFigFont#1#2#3#4#5{%
  \reset@font\fontsize{#1}{#2pt}%
  \fontfamily{#3}\fontseries{#4}\fontshape{#5}%
  \selectfont}%
\fi\endgroup%
\begin{picture}(4905,2580)(661,3824)
\put(1171,6239){\makebox(0,0)[b]{\smash{{\SetFigFont{12}{14.4}{\rmdefault}{\mddefault}{\updefault}{\color[rgb]{.56,0,0}$d_1$}%
}}}}
\put(2431,6239){\makebox(0,0)[b]{\smash{{\SetFigFont{12}{14.4}{\rmdefault}{\mddefault}{\updefault}{\color[rgb]{.56,0,0}$d_2$}%
}}}}
\put(2881,5339){\makebox(0,0)[b]{\smash{{\SetFigFont{12}{14.4}{\rmdefault}{\mddefault}{\updefault}{\color[rgb]{.56,0,0}$d_3$}%
}}}}
\put(721,5339){\makebox(0,0)[b]{\smash{{\SetFigFont{12}{14.4}{\rmdefault}{\mddefault}{\updefault}{\color[rgb]{.56,0,0}$d_5$}%
}}}}
\put(1891,5429){\makebox(0,0)[b]{\smash{{\SetFigFont{12}{14.4}{\rmdefault}{\mddefault}{\updefault}{\color[rgb]{.56,0,0}$d_4$}%
}}}}
\put(4051,5744){\makebox(0,0)[b]{\smash{{\SetFigFont{12}{14.4}{\rmdefault}{\mddefault}{\updefault}{\color[rgb]{.56,0,0}$\scriptstyle d_4$}%
}}}}
\put(5131,5339){\makebox(0,0)[b]{\smash{{\SetFigFont{12}{14.4}{\rmdefault}{\mddefault}{\updefault}{\color[rgb]{.56,0,0}$d_2$}%
}}}}
\put(3871,5339){\makebox(0,0)[b]{\smash{{\SetFigFont{12}{14.4}{\rmdefault}{\mddefault}{\updefault}{\color[rgb]{.56,0,0}$d_5$}%
}}}}
\put(5446,5879){\makebox(0,0)[b]{\smash{{\SetFigFont{12}{14.4}{\rmdefault}{\mddefault}{\updefault}{\color[rgb]{.56,0,0}$d_3$}%
}}}}
\put(3511,5879){\makebox(0,0)[b]{\smash{{\SetFigFont{12}{14.4}{\rmdefault}{\mddefault}{\updefault}{\color[rgb]{.56,0,0}$d_1$}%
}}}}
\put(1171,4889){\makebox(0,0)[b]{\smash{{\SetFigFont{12}{14.4}{\rmdefault}{\mddefault}{\updefault}{\color[rgb]{.56,0,0}$d_1$}%
}}}}
\put(2431,4889){\makebox(0,0)[b]{\smash{{\SetFigFont{12}{14.4}{\rmdefault}{\mddefault}{\updefault}{\color[rgb]{.56,0,0}$d_2$}%
}}}}
\put(2881,3989){\makebox(0,0)[b]{\smash{{\SetFigFont{12}{14.4}{\rmdefault}{\mddefault}{\updefault}{\color[rgb]{.56,0,0}$d_3$}%
}}}}
\put(721,3989){\makebox(0,0)[b]{\smash{{\SetFigFont{12}{14.4}{\rmdefault}{\mddefault}{\updefault}{\color[rgb]{.56,0,0}$d_5$}%
}}}}
\put(2026,4079){\makebox(0,0)[b]{\smash{{\SetFigFont{12}{14.4}{\rmdefault}{\mddefault}{\updefault}{\color[rgb]{.56,0,0}$d_4$}%
}}}}
\put(3511,4529){\makebox(0,0)[b]{\smash{{\SetFigFont{12}{14.4}{\rmdefault}{\mddefault}{\updefault}{\color[rgb]{.56,0,0}$d_1$}%
}}}}
\put(5446,4529){\makebox(0,0)[b]{\smash{{\SetFigFont{12}{14.4}{\rmdefault}{\mddefault}{\updefault}{\color[rgb]{.56,0,0}$d_2$}%
}}}}
\put(5131,3989){\makebox(0,0)[b]{\smash{{\SetFigFont{12}{14.4}{\rmdefault}{\mddefault}{\updefault}{\color[rgb]{.56,0,0}$d_3$}%
}}}}
\put(3871,3989){\makebox(0,0)[b]{\smash{{\SetFigFont{12}{14.4}{\rmdefault}{\mddefault}{\updefault}{\color[rgb]{.56,0,0}$d_5$}%
}}}}
\put(4501,4664){\makebox(0,0)[b]{\smash{{\SetFigFont{12}{14.4}{\rmdefault}{\mddefault}{\updefault}{\color[rgb]{.56,0,0}$d_4$}%
}}}}
\put(676,5789){\makebox(0,0)[b]{\smash{{\SetFigFont{12}{14.4}{\rmdefault}{\mddefault}{\updefault}{\color[rgb]{0,0,0}$T_1$:}%
}}}}
\put(676,4439){\makebox(0,0)[b]{\smash{{\SetFigFont{12}{14.4}{\rmdefault}{\mddefault}{\updefault}{\color[rgb]{0,0,0}$T_2$:}%
}}}}
\put(2926,5789){\makebox(0,0)[b]{\smash{{\SetFigFont{12}{14.4}{\rmdefault}{\mddefault}{\updefault}{\color[rgb]{0,0,0}$\Rightarrow$}%
}}}}
\put(2926,4439){\makebox(0,0)[b]{\smash{{\SetFigFont{12}{14.4}{\rmdefault}{\mddefault}{\updefault}{\color[rgb]{0,0,0}$\Rightarrow$}%
}}}}
\end{picture}%

%% file: Flatness.tex
\section{The Polytope of Compactifications}
\label{sec:compactification}

By definition, the top describes the degeneration of the elliptic
fiber over a toric divisor in the base. The base divisor is one of the
rays in the base fan. The obvious question is how this data can be
completed to a compact Calabi-Yau manifold, that is, how to complete
the top and the choice of base fan to a reflexive polytope. In fact,
this has a nice answer: The remaining choices for a lattice polytope
after fixing the tops and the base again amount to the integral points
of a polytope.\footnote{This polytope is not necessarily integral,
  that is, its vertices are in general rational.} This just follows
from convexity, and one needs to verify reflexivity and flatness of
the fibration by hand.

In particular, we will be interested in the case of a single $SU(5)$
top together with trivial tops over the remaining rays of the base
fan. For the purposes of this section, we will only consider the case
where the base fan equals $\CP^n$, whose rays are generated by the
unit vectors $e_1$, $\dots$, $e_n$ together with $-\sum e_i$. Then
\begin{itemize}
\item The fixed $SU(5)$ top can be chosen to project to $[0, e_1]$.
\item The single point generating the trivial top over each of $e_2$,
  $\dots$, $e_n$ can be chosen to have fiber coordinates $(0,0)$ by a
  $GL(n,\Z)$ rotation fixing all previous tops.
\item The final point, generating the trivial top over $-\sum e_i$,
  has coordinates $(p_1, p_2) \in \Z^2$ with no remaining freedom of
  coordinate redefinition.
\end{itemize}
This parametrizes the choices of completion to a polytope by a pair of
integers $(p_1, p_2)$. These are constrained by convexity: Having
fixed the height-one points of the other tops, there is only a finite
range of $(p_1, p_2)$ such that the fiber (preimage of the origin in
the base) of the convex hull does not exceed the chosen fiber
polygon. These are linear constraints, turning the allowed region for
$(p_1, p_2)$ into a polygon (with not necessarily integral
vertices). 

It turns out that all lattice polytopes for a single $SU(5)$ top over
$\CP^n$ that one constructs just by demanding convexity, as above, are
automatically reflexive. Their total number for small values of $n$ is
listed in \autoref{tab:reflexivecount}. We included also the cases
$\mathbb{P}^4$ and $\mathbb{P}^5$ that, when used as base, would not
lead to gauge theories in four dimensions. However, the construction
can be supplemented by additional polynomials specifying the actual
base as hypersurface in $\mathbb{P}^4$ or complete intersection in
$\mathbb{P}^5$. For example, the Fano threefold obtained by a quartic
constraint in $\mathbb{P}^4$ is a viable choice for the base. Note
that realizing the base itself as hypersurface or complete
intersection can be also phenomenologically motivated. Such
realizations allow for more exhaustive choices of fluxes on the GUT
brane as demonstrated in the models of~\cite{Blumenhagen:2009yv,
  Grimm:2009yu}.  This applies in particular to hypercharge
flux~\cite{Beasley:2008kw, Donagi:2008kj, Palti:2012dd} that is
non-trivial on the GUT brane but trivial on the entire base manifold.
Our construction thus extends straightforwardly to these more involved
Calabi-Yau fourfold examples.
\begin{table}
  \centering
  \begin{tabular}{ccccccc}
    \toprule
    Fiber & Top & 
    $N^{SU(5)}_{\CP^1}$ &
    $N^{SU(5)}_{\CP^2}$ &
    $N^{SU(5)}_{\CP^3}$ &
    $N^{SU(5)}_{\CP^4}$ &
    $N^{SU(5)}_{\CP^5}$
    \\ \midrule
    $F_1$ & $\tau_{ 1,1 }$ & $1$ & $5$ & $12$ & $22$ & $35$ \\ 
    $F_1$ & $\tau_{ 1,2 }$ & $1$ & $5$ & $12$ & $22$ & $35$ \\ 
    $F_1$ & $\tau_{ 1,3 }$ & $1$ & $4$ & $8$ & $14$ & $21$ \\ 
    $F_1$ & $\tau_{ 1,4 }$ & $1$ & $5$ & $12$ & $22$ & $35$ \\ 
    $F_1$ & $\tau_{ 1,5 }$ & $1$ & $5$ & $11$ & $18$ & $27$ \\ 
    $F_2$ & $\tau_{ 2,1 }$ & $2$ & $9$ & $20$ & $30$ & $42$ \\ 
    $F_2$ & $\tau_{ 2,2 }$ & $3$ & $10$ & $21$ & $36$ & $55$ \\ 
    $F_2$ & $\tau_{ 2,3 }$ & $3$ & $8$ & $15$ & $24$ & $35$ \\ 
    $F_3$ & $\tau_{ 3,1 }$ & $2$ & $9$ & $20$ & $35$ & $54$ \\ 
    $F_3$ & $\tau_{ 3,2 }$ & $3$ & $10$ & $21$ & $36$ & $55$ \\ 
    $F_3$ & $\tau_{ 3,3 }$ & $3$ & $10$ & $21$ & $36$ & $55$ \\ 
    $F_3$ & $\tau_{ 3,4 }$ & $3$ & $10$ & $21$ & $36$ & $55$ \\ 
    $F_3$ & $\tau_{ 3,5 }$ & $3$ & $10$ & $21$ & $36$ & $55$ \\ 
    $F_3$ & $\tau_{ 3,6 }$ & $3$ & $10$ & $21$ & $36$ & $55$ \\ 
    $F_3$ & $\tau_{ 3,7 }$ & $3$ & $10$ & $21$ & $36$ & $55$ \\ 
    $F_4$ & $\tau_{ 4,1 }$ & $3$ & $10$ & $21$ & $36$ & $55$ \\ 
    $F_4$ & $\tau_{ 4,2 }$ & $3$ & $10$ & $21$ & $36$ & $55$ \\ 
    $F_4$ & $\tau_{ 4,3 }$ & $3$ & $10$ & $21$ & $36$ & $55$ \\ 
    $F_5$ & $\tau_{ 5,1 }$ & $6$ & $12$ & $20$ & $31$ & $44$ \\ 
    $F_5$ & $\tau_{ 5,2 }$ & $5$ & $15$ & $30$ & $50$ & $75$ \\ 
    $F_5$ & $\tau_{ 5,3 }$ & $5$ & $15$ & $30$ & $50$ & $75$ \\ 
    $F_5$ & $\tau_{ 5,4 }$ & $6$ & $16$ & $31$ & $51$ & $76$ \\ 
    $F_5$ & $\tau_{ 5,5 }$ & $5$ & $15$ & $30$ & $50$ & $75$ \\ 
    $F_6$ & $\tau_{ 6,1 }$ & $6$ & $16$ & $31$ & $51$ & $76$ \\ 
    $F_6$ & $\tau_{ 6,2 }$ & $6$ & $16$ & $31$ & $51$ & $76$ \\ 
    $F_6$ & $\tau_{ 6,3 }$ & $5$ & $15$ & $30$ & $50$ & $75$ \\ 
    $F_6$ & $\tau_{ 6,4 }$ & $5$ & $15$ & $30$ & $50$ & $75$ \\ 
    $F_6$ & $\tau_{ 6,5 }$ & $6$ & $16$ & $31$ & $51$ & $76$ \\ 
    $F_7$ & $\tau_{ 7,1 }$ & $8$ & $18$ & $30$ & $45$ & $63$ \\ 
    $F_8$ & $\tau_{ 8,1 }$ & $8$ & $21$ & $40$ & $65$ & $96$ \\ 
    $F_8$ & $\tau_{ 8,2 }$ & $8$ & $21$ & $40$ & $65$ & $96$ \\ 
    $F_9$ & $\tau_{ 9,1 }$ & $8$ & $21$ & $40$ & $65$ & $96$ \\ 
    $F_9$ & $\tau_{ 9,2 }$ & $8$ & $21$ & $40$ & $65$ & $96$ \\ 
    $F_{10}$ & $\tau_{ 10,1 }$ & $8$ & $21$ & $40$ & $65$ & $96$ \\ 
    $F_{11}$ & $\tau_{ 11,1 }$ & $11$ & $27$ & $50$ & $80$ & $117$ \\ 
    $F_{12}$ & $\tau_{ 12,1 }$ & $11$ & $27$ & $50$ & $80$ & $117$ \\ 
    $F_{14}$ & $\tau_{ 14,1 }$ & $14$ & $23$ & $38$ & $57$ & $80$ \\ 
    \bottomrule
  \end{tabular}
  \caption[Number of $SU(5)$ models over $\CP^n$]{Number
    $N^{SU(5)}_{\CP^n}$ of reflexive polytopes fibered over $\CP^n$
    with one $SU(5)$-top and $n$ trivial tops, modulo fiber-preserving automorphisms.}
  \label{tab:reflexivecount}
\end{table}

\section{Flatness of the Fibration}
\label{sec:flatness}

Not all compactifications of F-theory give rise to ordinary gauge theories, as
they may contain tensionless strings yielding an infinite tower of
massless fields in the low-energy effective action. While there is
nothing wrong with that, these theories have to be excluded when one
looks for phenomenologically viable theories. Alternatively, one can
try to lift all but finitely many of these massless fields through
fluxes, but we will leave this for future work. The geometric origin
of these massless strings \cite{Seiberg:1996vs, Morrison:1996na,
  Morrison:1996pp, Candelas:2000nc} are three-branes wrapping a curve
inside a surface of vanishing volume in the F-theory limit. Such a surface
must necessarily sit over a point in the discriminant locus, that is,
in a fiber of the elliptic fibration that is at least two-dimensional.
Clearly, this cannot happen if all degenerate fibers are of
Kodaira type. Hence, any $K3$ hypersurface in a toric variety
constructed by gluing two tops along the fiber polygon has all fibers
one-dimensional. A fibration with the property that all fibers are of
the same dimension is called \emph{flat}.\footnote{Flat in the sense
  of homological algebra, that is, the functions in a neighborhood of
  each fiber are a flat module over the function ring of the base.}

For the case of hypersurfaces in toric varieties, there are two
possible sources for non-flat fibers:
\begin{itemize}
\item The ambient toric fiber can jump in dimension. That is, the
  toric fibration of the ambient space can already fail to be
  flat~\cite{Braun:2011ux}. This happens in particular if one places
  two non-Abelian tops on neighboring base rays such that the
  intersection is not a Miranda model~\cite{MR690264}.
\item Even if the ambient toric fibration is flat, the hypersurface
  equation can vanish identically in the fiber direction for certain
  fibers. Then the fiber of the elliptic fibration becomes
  two-dimensional.
\end{itemize}
The flatness of the ambient toric fibration can easily be
checked~\cite{2000math.....10082H, BraunNovoseltsev:toric_variety,
  Braun:2011ux} using toric methods. In particular, this is always the
case when only a single non-trivial top is used. Hence, we will focus
in the remainder of this paper on the second source for non-flat
fibers. In this case, the non-flat fibers do not generally lie over toric
fixed points.

\subsection{Codimension Two Fibers}
\label{sec:flat2}

While elliptic $K3$s are always flat fibrations, a toric Calabi-Yau
threefold hypersurface can be non-flat even if the ambient toric
fibration is. These codimension-two (over the base, codimension one
inside the discriminant) non-flat fibers come from lattice points
interior to the $z=1$ facets. This is because a point interior to a
facet corresponds to a toric divisor such that the hypersurface
equation restricts to a (generically non-zero) constant. However, a
point interior to a facet of the top is usually not interior to a
facet of the entire $4$-d polytope. This means that the hypersurface
equation is \emph{not} constant on the corresponding divisor in the
ambient space, but only in the fiber direction. In fact, this fiber-wise
constant is a section of a nef line bundle over the (toric)
discriminant component, and therefore has a zero somewhere. This is
the location of the non-flat fiber.

There is one loophole in the argument: If the base ray, over which the
non-trivial top is placed, is itself a point interior to a facet of
the base polytope, then a point interior to a facet of the top is also
interior to a facet of the $4$-d polytope. Geometrically, this means
that the discriminant component is a curve of self-intersection $-2$
and the hypersurface again avoids the corresponding toric divisor
entirely. However, this is not a physically desirable situation: The
hypersurface equation restricted to this discriminant component is now
independent of the point along the discriminant. Therefore, there are no
codimension-two degenerations at all, and in particular no matter
curves. Hence we will not consider this case in the following, and
only allow tops with no points interior to facets.

For example, consider the del Pezzo surface of degree $7$, that is
$F_5$ in \autoref{fig:tops}, as the fiber polygon. Then one of the
tops, namely $\tau_{5,1}$, will have non-flat fibers and the remaining
$4$ tops $\tau_{5,2}$, $\dots$, $\tau_{5,5}$ yield flat fibrations in
codimension two.

\subsection{Flatness Criterion}
\label{sec:criterion}

Having described the flatness criterion for codimension-two fibers, we
now proceed to generalize it to arbitrary codimension. As an example,
we then apply it to the physically relevant case of codimension-three
fibers in elliptically fibered fourfolds.

By an analogous argument as in the previous section, a Calabi-Yau
hypersurface in an ambient flat fibration will be flat itself if the
hypersurface equation never vanishes identically in the fiber
direction. For simplicity, consider the case where there is only a
single non-trivial top. To understand the hypersurface equation we
collect the monomials of the hypersurface equation $p=0$ by their
dependence on the top homogeneous coordinates $z_\tau = \{z_{\tau,1},
\dots, z_{\tau,k}\}$ as
\begin{equation}
  p(z_\tau, z) = 
  \sum_{\vec{\imath}=(i_1,\dots,i_k)\in I} 
  z_\tau^{\vec{\imath}}
  ~p_{\vec{\imath}}(z)
  =
  \sum_{\vec{\imath}\in I} z_\tau^{\vec{\imath}}
  \Big(
    \sum_{\vec{\jmath}\in J_{\vec{\imath}}} 
    a_{\vec{\imath}\;\vec{\jmath}} z^{\vec{\jmath}}
  \Big)
  ,\qquad
  a_{\vec{\imath}\;\vec{\jmath}} \in \C.
\end{equation}
The irreducible components of the degenerate fiber induced by the top
are the toric divisors $z_{\tau,\ell}=0$ corresponding to the integral
points of the top that are not in the fiber polygon. One needs to
check for every irreducible component that the fibration is flat. The
irreducible fiber component $z_{\tau,\ell}=0$ projects to one
discriminant component $D_\tau$, and the local coordinates on $D_\tau$
are the rays in the star of $\pi(\tau)$ in the base. Each of the
polynomials $p_{\vec{\imath}}(z)$ only depends on the base coordinates
and therefore defines a divisor $V_\tau(p_{\vec{\imath}}) \subset
D_\tau$ on the discriminant. Then a generic hypersurface is a flat
fibration over $D_\tau$ if these divisors do not meet, that is,
\begin{equation}
  D_\tau \owns
  \bigcap_{
    \begin{smallmatrix}
      \vec{\imath} \in I \\
      z_\tau^\ell \nmid z_\tau^{\vec{\imath}}
    \end{smallmatrix}
  }
  V_\tau(p_{\vec{\imath}})
  = 
  \bigcap_{
    \begin{smallmatrix}
      {\vec{\imath}=(i_1,\dots,i_k)\in I} \\
      i_{\ell} = 0
    \end{smallmatrix}
  }
  V_\tau(p_{\vec{\imath}})
  = 
  \emptyset.
\end{equation}
The summation range $I$ is over all fiber monomials, that is, integral
points of the dual of the top polytope intersected with the projection
of the dual polytope of the ambient toric variety. The summation range
$J_{\vec{\imath}}$ is the fiber of the projection of the dual
polytope, that is, over all integral points of the dual polytope whose
monomial is divisible by $z_\tau^{\vec{\imath}}$.

Phrasing \autoref{sec:flat2} in this language, if $z_{\tau,\ell}$
corresponds to the ray generated by an integral point interior to a
facet of the top then $I=\{\vec{\imath}\}$ consists of only a single
element. The corresponding divisor $V_\tau(p_{\vec{\imath}})\subset
D_\tau$ will generically be non-empty and, therefore, the fibration
non-flat. The only loophole is if the divisor \emph{is} empty, that
is, $J_{\vec{\imath}} = \{\vec{\jmath}\}$ consists of a single point
which then must be a vertex of the dual polytope. But this means that
the point was not just interior to a facet of the top, but interior to
a facet (dual to $\vec{\jmath}$) of the ambient toric variety.

\subsection{Codimension Three Fibers}
\label{sec:flat3}

We now apply the flatness criterion to Calabi-Yau fourfold
hypersurfaces. As we will see, flatness is a non-generic property in
the sense that it imposes additional equations on the polytope of
compactifications (as defined in
\autoref{sec:compactification}). Hence, the flat fourfold fibrations
are identified with integral points in a strictly
smaller-dimensional polytope than the set of all convex lattice
polytopes with the specified top and base.

The new source for non-flat fibers are irreducible fiber components
such that there are only two distinct fiber monomials. These are
integral points of the top such that their dual face in the dual top
contains exactly two points, that is, such that the dual face is an
interval. In other words, the corresponding integral point of the top
is along an edge of the top such that it is contained in only two
$2$-faces. Note that this is the case for \emph{every} $SU(5)$-top
that is not already non-flat in codimension $2$ due to an integral
point interior to a facet. This is because the polygon at height $z=1$
of the $SU(5)$-top, see \autoref{fig:tops}, is a lattice polygon with
circumference $5$ in lattice units. But such a lattice polygon has
either an interior point or is degenerate, see also
\autoref{fig:tops}. Therefore, each $SU(5)$ top that is flat in
codimension-two yields a non-trivial flatness condition in
codimension-three associated to the integral point on the edge.

For simplicity, let us assume that the discriminant component $D_\tau$
of the $SU(5)$-top is a toric surface where any two effective divisors
intersect. This will always be the case in the examples below, where
we will be using $D_\tau=\CP^2$. Consider now the toric divisor
$z_{\tau,\ell}=0$ corresponding to the integral point interior to an
edge. The index set $I=\{\vec{\imath}^{(0)}, \vec{\imath}^{(1)}\}$
consists of two elements, corresponding to the two facets
$F_{\tau,2}^{(0)}$, $F_{\tau,2}^{(1)}$ of the top adjacent to the
edge. The fibration is then flat if and only if one of the divisors is
trivial, say, $V_\tau(p_{\vec{\imath}^{(1)}})\subset D_\tau$. This is
the case if $J_{\vec{\imath}^{(1)}}$ contains a single element, which
then must be a vertex of the dual ambient space polytope. Hence, the
facet $F_{\tau,2}^{\vec{\imath}^{(1)}}$ of the top is contained in
only a single facet of the ambient space polytope. Note that one of
the facets $F_{\tau,2}^{(0)}$, $F_{\tau,2}^{(1)}$ of the $SU(5)$-top
will be parallel to the fiber polygon and the other will contain at
least one point of the fiber polygon. The former will always be
contained in at least two facets of the ambient space unless the base
ray $\pi(\tau)$ is an interior point of a facet of the base
polytope. As discussed in \autoref{sec:flat2}, this is not a
particularly interesting case and we will ignore it in the
following. Therefore, the facet $F_{\tau,2}^{\vec{\imath}^{(1)}}$ of
interest is the one that contains at least one point of the fiber
polygon.

\subsection{Examples}
\label{sec:example}

The constraints from flatness of the fibration can rule out a fixed
combination of top and base polytope. To see this explicitly, we will
look at two examples in this section, namely the top $\tau_{10,1}$ and
$\tau_{3,6}$, respectively, to construct an elliptic fibration over
$\CP^3$. Note that $\tau_{10,1}$ is the unique $SU(5)$-top in
Weierstrass form, that is, with ambient space fiber
$\CP^2[1,2,3]$. The $\tau_{3,6}$ top was used in~\cite{Braun:2013yti}
and we will follow the same coordinate choice, which is different from
\autoref{fig:tops}, but of course $GL(2,\Z)$-equivalent to it.
\begin{table}
  \centering
  \begin{minipage}{0.45\linewidth}
    \begin{tabular}{|c||cc|ccc|}
      \hline
      $\nabla_{10,1}(p_1, p_2)$ & 
      \multicolumn{2}{c|}{fiber} & 
      \multicolumn{3}{c|}{base}
      \\ \hline \hline
      \multirow{4}{*}{fiber}
      & 1  &  0 & 0 & 0 & 0 \\
      & 0  &  1 & 0 & 0 & 0 \\
      & -3 & -2 & 0 & 0 & 0 \\ \hline
      \multirow{5}{*}{$\tau_{10,1}$}
      & 0  &  0 & 1 & 0 & 0 \\
      &-1  &  0 & 1 & 0 & 0 \\
      &-2  & -1 & 1 & 0 & 0 \\
      &-3  & -2 & 1 & 0 & 0 \\
      &-1  & -1 & 1 & 0 & 0 \\\hline
      \multirow{1}{*}{trivial top}
      & 0  &  0 & 0 & 1 & 0 \\  \hline
      \multirow{1}{*}{trivial top}
      & 0  &  0 & 0 & 0 & 1 \\  \hline
      \multirow{1}{*}{trivial top}
      & $p_1$  &  $p_2$ &-1 &-1 &-1 \\  \hline
    \end{tabular}
  \end{minipage}
  \begin{minipage}{0.45\linewidth}
    \begin{tabular}{|c||cc|ccc|}
      \hline
      $\nabla_{3,6}(p_1, p_2)$ & 
      \multicolumn{2}{c|}{fiber} & 
      \multicolumn{3}{c|}{base}
      \\ \hline \hline
      \multirow{4}{*}{fiber}
      &  1 &  0 & 0 & 0 & 0 \\ 
      &  0 &  1 & 0 & 0 & 0 \\
      & -1 &  0 & 0 & 0 & 0 \\ 
      & -1 & -1 & 0 & 0 & 0 \\  \hline
      \multirow{5}{*}{$\tau_{3,6}$}
      & -2 & -1 & 1 & 0 & 0 \\ 
      & -1 & -1 & 1 & 0 & 0 \\
      & -1 &  0 & 1 & 0 & 0 \\ 
      &  0 & -1 & 1 & 0 & 0 \\
      &  0 &  0 & 1 & 0 & 0 \\  \hline
      \multirow{1}{*}{trivial top}
      &  0 &  0 & 0 & 1 & 0 \\ \hline 
      \multirow{1}{*}{trivial top}
      &  0 &  0 & 0 & 0 & 1 \\  \hline
      \multirow{1}{*}{trivial top}
      & $p_1$  &  $p_2$ &-1 &-1 &-1 \\  \hline
    \end{tabular}
  \end{minipage}
  \caption[Parametrization of all polytopes with base $\CP^3$ and top
  $\tau_{10,1}$ and $\tau_{3,6}$]{
    Parametrization $(p_1,p_2)\in \Z^2$ 
    of all polytopes with base $\CP^3$ and top
    $\tau_{10,1}$ (left) and $\tau_{3,6}$ (right), respectively. The
    fibration is the projection on the last 3 coordinates.}
  \label{tab:example}
\end{table}
As described in \autoref{sec:compactification}, we can choose
coordinates such that everything except the fiber coordinates of a
single point are fixed. These are shown in \autoref{tab:example}.

Imposing convexity of the $5$-d polytopes amounts to the inequalities
\begin{equation}
  \begin{split}
    \nabla_{10,1}(p_1, p_2):\quad &
    p_1 + p_2 \leq 4
    ,\quad
    -p_1 + p_2  \leq 3
    ,\quad
    p_1 - 2 p_2 \leq 3
    \\
    \nabla_{3,6}(p_1, p_2):\quad &
    p_1 + p_2 \leq 4
    ,\quad
    -p_1 \leq 2
    ,\quad
    p_1 - 2 p_2 \leq 2
    ,\quad
    - p_1 + p_2 \leq 3
    .
  \end{split}
  \label{eq:convex}
\end{equation}
The interior point of an edge in the $\tau_{10,1}$-top is $(-2, -1,
1,0,0)$. The relevant $2$-face of the top for the flatness criterion
is
\begin{equation}
  F_{\tau_{10,1},2}^{\vec{\imath}^{(1)}}   =
  \big\langle         
  (-3, -2, 1, 0, 0),
  (-1,  0, 1, 0, 0),
  (-3, -2, 0, 0, 0),
  ( 0,  1, 0, 0, 0)
  \big\rangle
\end{equation}
This facet is contained in a the $(p_1, p_2)$-independent supporting
hyperplane of the total polytope $\nabla_{10,1}(p_1, p_2)$ defined by
\begin{equation}
  (1, -1, 0, -1, -1) \cdot \vec{x} + 1 = 0\,.
\end{equation}
It is contained in further facets
of $\nabla_{10,1}$ unless the final point $(p_1, p_2, -1, -1, -1)$ is
also on this hyperplane, and therefore cannot span an independent
facet. This is a linear equation for $(p_1, p_2)$. Together with the
result for the second example, this equation is
\begin{equation}
  \begin{split}
    \nabla_{10,1}(p_1, p_2):\quad &
    p_1 - p_2 = -3
    \\
    \nabla_{3,6}(p_1, p_2):\quad &
    p_2 = 3
  \end{split}
  \label{eq:flat}
\end{equation}
\begin{figure}
  \centering
  \includegraphics[width=0.45\linewidth]{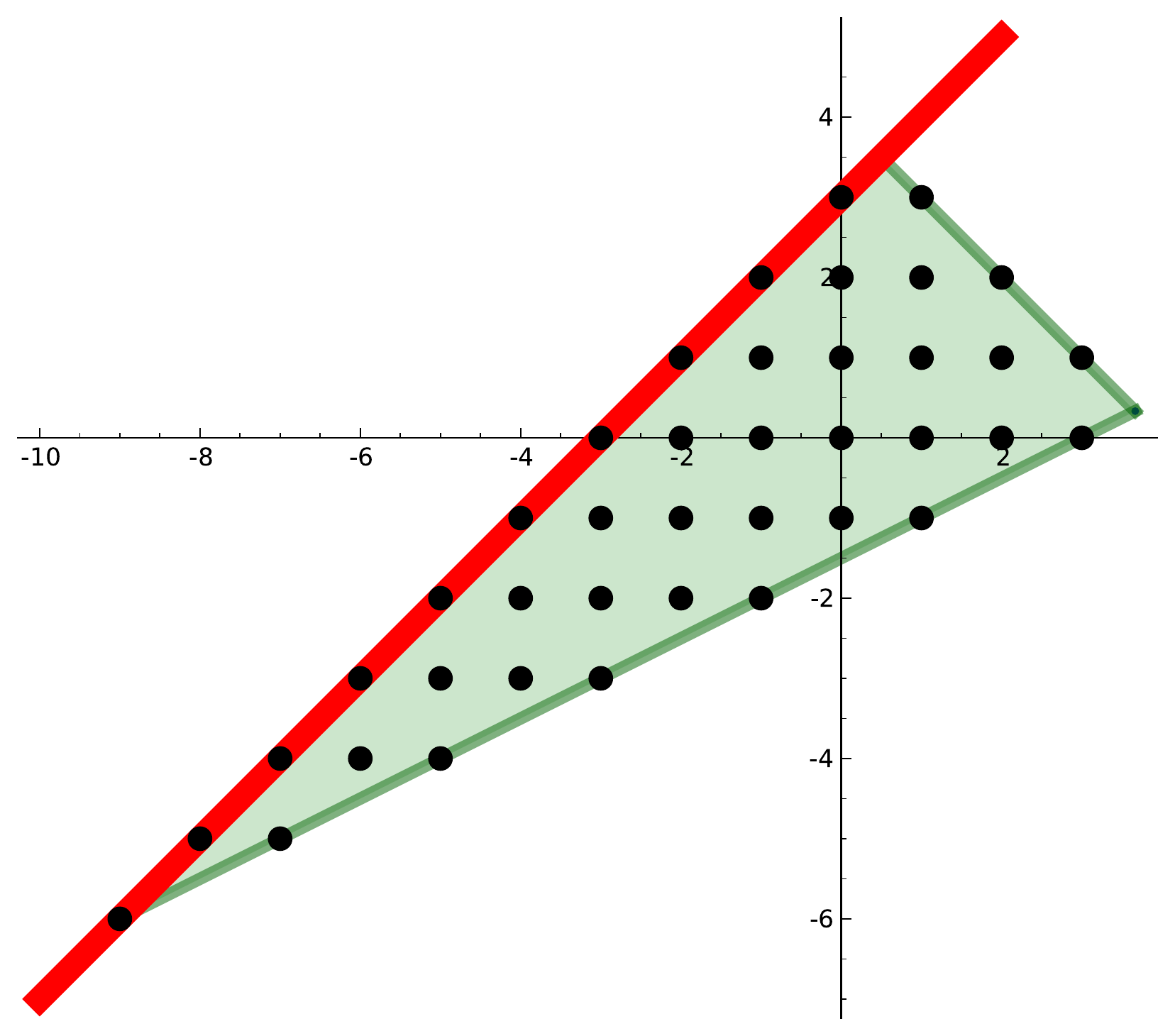}
  \includegraphics[width=0.45\linewidth]{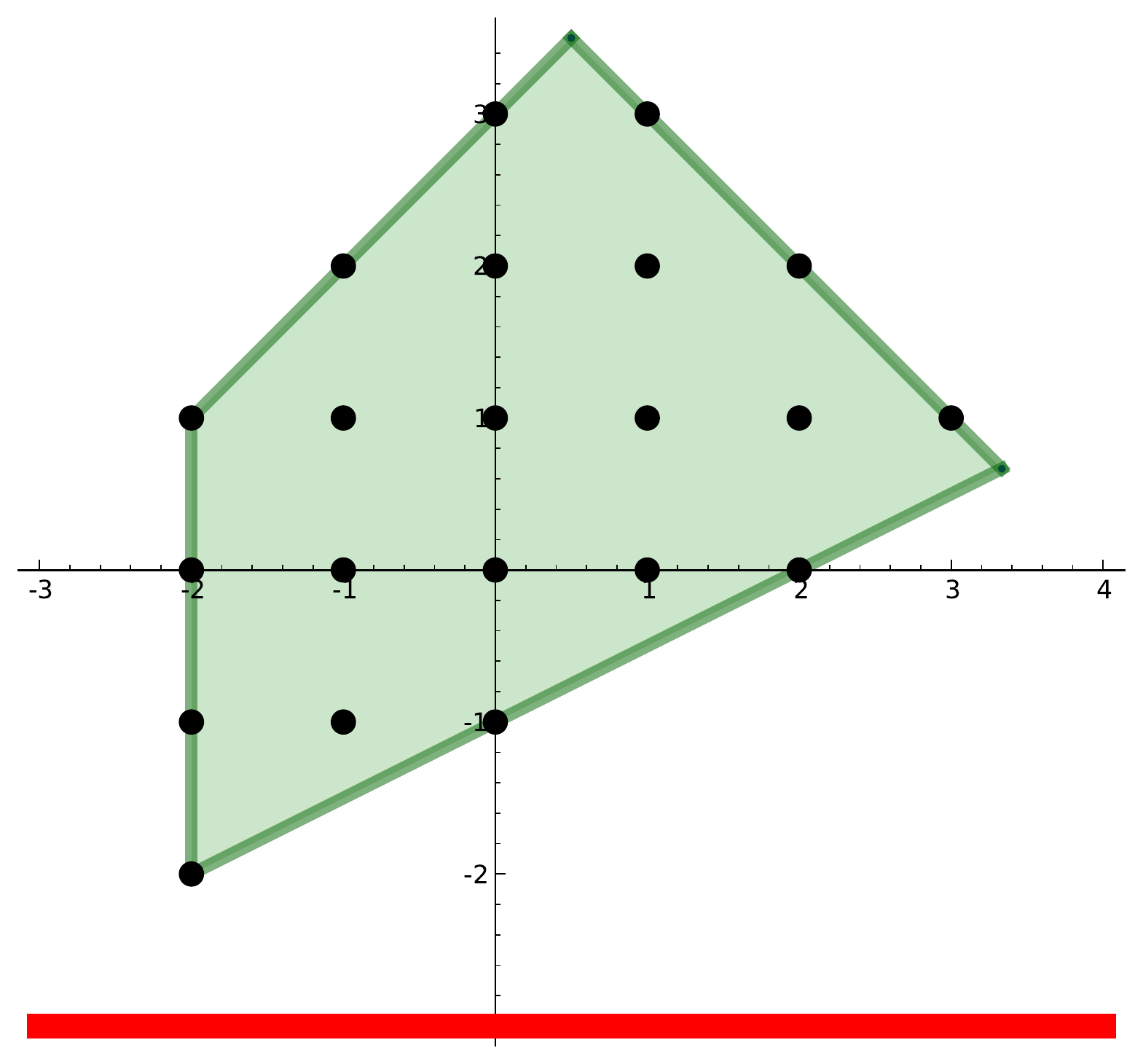}
  \caption[$SU(5)$ models with top $\tau_{10,1}$ and $\tau_{3,6}$]{The
    black points are the solution set $(p_1, p_2)$ for $SU(5)$ models
    with top $\tau_{10,1}$ (left) and $\tau_{3,6}$ (right) fibered
    over $\CP^3$. The green polygon is the convexity constraint from
    eq.~\eqref{eq:convex}. The red line is the condition of flatness
    of the fibration, see eq.~\eqref{eq:flat}.}
  \label{fig:example}
\end{figure}
The constraints coming from convexity and flatness are shown in
\autoref{fig:example}. We observe that there are many flat elliptic
fibrations using the $\tau_{10,1}$ top, but none with the $\tau_{3,6}$
top.

\subsection{Flattening Base Change}
\label{sec:flatten}

It is perhaps unexpected that for $\CP^3$, the simplest choice of base
for a Calabi-Yau fourfold, the top $\tau_{3,6}$ cannot be used to
construct a flat elliptic fibration. One might speculate that this has
something to do with the non-standard matter $U(1)$-charges that were
found in~\cite{Braun:2013yti}. However, this is not the case and one
can easily build flat elliptic fibrations over other bases, leading to
$4$-d $SU(5)$ GUT gauge theories with matter charges among
$\Rep{5}_{-2}$, $\Rep{5}_3$, $\Rep{5}_7$, $\Rep{5}_8$, $\Rep{10}_1$,
and conjugates. In fact, already the base $\CP^1\times \CP^2$ works
for this purpose. Since it is instructional to consider a different
base than just $\CP^n$, we will give some of the details of the
possible reflexive polytopes for this base manifold.
\begin{table}
  \centering
  \begin{tabular}{|c||cc|ccc|}
    \hline
    $\nabla_{3,6}^{\CP^1\times\CP^2}(p_1, p_2; p_3, p_4)$ & 
    \multicolumn{2}{c|}{fiber} & 
    \multicolumn{3}{c|}{base}
    \\ \hline \hline
    \multirow{4}{*}{fiber}
    &  1 &  0 & 0 & 0 & 0 \\ 
    &  0 &  1 & 0 & 0 & 0 \\
    & -1 &  0 & 0 & 0 & 0 \\ 
    & -1 & -1 & 0 & 0 & 0 \\  \hline
    \multirow{5}{*}{$\tau_{3,6}$}
    & -2 & -1 & 1 & 0 & 0 \\ 
    & -1 & -1 & 1 & 0 & 0 \\
    & -1 &  0 & 1 & 0 & 0 \\ 
    &  0 & -1 & 1 & 0 & 0 \\
    &  0 &  0 & 1 & 0 & 0 \\  \hline
    \multirow{1}{*}{trivial top}
    &  $p_1$ &  $p_2$ & -1 & 0 & 0 \\ \hline 
    \multirow{1}{*}{trivial top}
    &  0 &  0 & 0 & 1 & 0 \\ \hline 
    \multirow{1}{*}{trivial top}
    &  0 &  0 & 0 & 0 & 1 \\  \hline
    \multirow{1}{*}{trivial top}
    & $p_3$  &  $p_4$ &-1 &-1 &-1 \\  \hline
  \end{tabular}
  \caption[Parametrization of all polytopes with base
  $\CP^1\times\CP^2$ and top $\tau_{3,6}$]{
    Parametrization $(p_1,p_2;p_3,p_4)\in \Z^4$ 
    of all polytopes with base $\CP^1\times\CP^2$ and single $SU(5)$-top
    $\tau_{3,6}$ over $\ptset\times \CP^2$. The
    fibration is the projection on the last 3 coordinates.}
  \label{tab:flatP1xP2}
\end{table}

First of all, not all divisors of the base are equivalent any
more. For definiteness, we put the GUT divisor at $D_\tau = \ptset
\times \CP^2 \subset \CP^1\times \CP^2$. Up to coordinate changes,
there are now four integers parametrizing the possible embeddings in a
5-d polytope, see \autoref{tab:flatP1xP2}. The polytope of
compactifications is now 4-dimensional, and contains $75$ integral
points. These are the $75$ solutions to the convexity
constraints. Again, it turns out that for this choice of base all
polytopes that are allowed by convexity are actually reflexive. All
have $h^{1,1}=8$, corresponding to a single $U(1)$. Out of these, $3$
polytopes yield a flat fibration. These are
\begin{equation}
  (p_1,p_2;p_3,p_4)
  \in \big\{
  (0,0;-3,-3),~
  (0,1;-3,-3),~
  (1,1;-3,-3)
  \big\}
\end{equation}
Using the methods of \cite{Braun:2013yti}, one readily verifies that the non-Abelian
matter content once again consists of fields in the representations
$\Rep{5}_{-2}$, $\Rep{5}_3$, $\Rep{5}_7$, $\Rep{5}_8$, $\Rep{10}_1$ of $SU(5) \times U(1)$.
We hence conclude that it is easy to build $4$-d $SU(5)$-GUT models with the
$\tau_{3,6}$-top as long as the base is not $\CP^3$.


%% file: Conclusions.tex
\section{Conclusions}

We introduced an algorithm to construct compact elliptically fibered
Calabi-Yau fourfolds with specified gauge group and base space using
toric geometry. This provides a systematic approach to geometrically
engineering four-dimensional $\mathcal{N}=1$ gauge theories coupled to
gravity.  While our approach is otherwise general, we restricted
ourselves to situations in which the elliptic fiber is realized as a
hypersurface inside a toric ambient space. To exemplify the steps
involved, we constructed GUTs with non-Abelian gauge group $SU(5)$
while leaving the Abelian gauge sector unspecified.  We described
which parts of the construction can be performed independently of the
base space and classified all fibers and $SU(5)$ tops. It is crucial
to point out that a complete analysis depends on the entire Calabi-Yau
space.  First, the total number of $U(1)$ symmetries depends on the
base and its embedding in the fibration, and we introduced an
auxiliary polytope labeling all inequivalent choices for the
latter. Second, we showed that ensuring flatness of the fibration
imposes further non-trivial constraints that have to be checked for
each base manifold individually. We formulated the flatness
constraints as constraints on the toric data specifying the Calabi-Yau
manifold. With these techniques at hand, we constructed numerous
interesting fourfold examples with $SU(5) \times U(1)^n$ gauge group.

Using the methods from this paper, one is now in a position to perform
a systematic scan over all elliptically fibered Calabi-Yau fourfolds
embedded in toric ambient spaces over a fixed base.  Only restricting
the search to reflexive polytopes giving rise to a fixed gauge group
might turn this into a tractable problem with today's computer power.
One can then perform a survey of the different Abelian gauge
factors. While we made progress towards the combinatorial geometry
determining the matter spectrum and its $U(1)$ charges, there still
remains work to be done. In particular, it would be desirable to
determine all occurring matter representations from toric data alone,
without needing to specify a particular hypersurface equation.
Furthermore, it would be interesting to systematically study elliptic
fibrations with fibers realized as complete intersections in
higher-dimensional ambient spaces. This is in particular motivated by
phenomenology, since we presented a no-go theorem ruling out different
$U(1)$ charges for the antisymmetric representation of $SU(5)$.  We
therefore hope to return to this case in the
future~\cite{workinprogress}.

Having developed a formalism to extract the matter representations from
a compact geometry, one has to include $G$-fluxes to obtain a chiral
four-dimensional matter spectrum next.  This can be done
systematically by evaluating the three-dimensional Chern-Simons terms
as in~\cite{Grimm:2011fx, Cvetic:2012xn, Braun:2013yti}.  Doing so
requires a detailed knowledge of allowed triangulations of the toric
ambient spaces and, hence, of the phase structure in the
three-dimensional gauge theory~\cite{Grimm:2011fx, Hayashi:2013lra}.
Specifying the $G$-flux data allows one to completely determine the
matter content, which is localized at codimension two in the base. At
codimension three, a systematic analysis will be even more involved.
Nevertheless, there is hope that model building using the billions of
known Calabi-Yau fourfolds may soon become a tractable combinatorial
problem, which would allow to extract generic features or general
constraints of gauge theories coupled to gravity in F-theory.


%% file: EllipticCurves.tex
\section{The Group Law on a Cubic}
\label{a:curves}

In this short appendix let us briefly review the geometric origin of
the group law on the set of rational points of an elliptic curve.  For
a more exhaustive treatment of the extensive theory of elliptic curves
we refer to the literature, e.g.~\cite{Deligne1975courbes}.  We begin
by discussing the group law in the case of the Weierstrass cubic. This
is the affine curve
\begin{figure}[htb]
  \centering
  \includegraphics[width=0.7\textwidth]{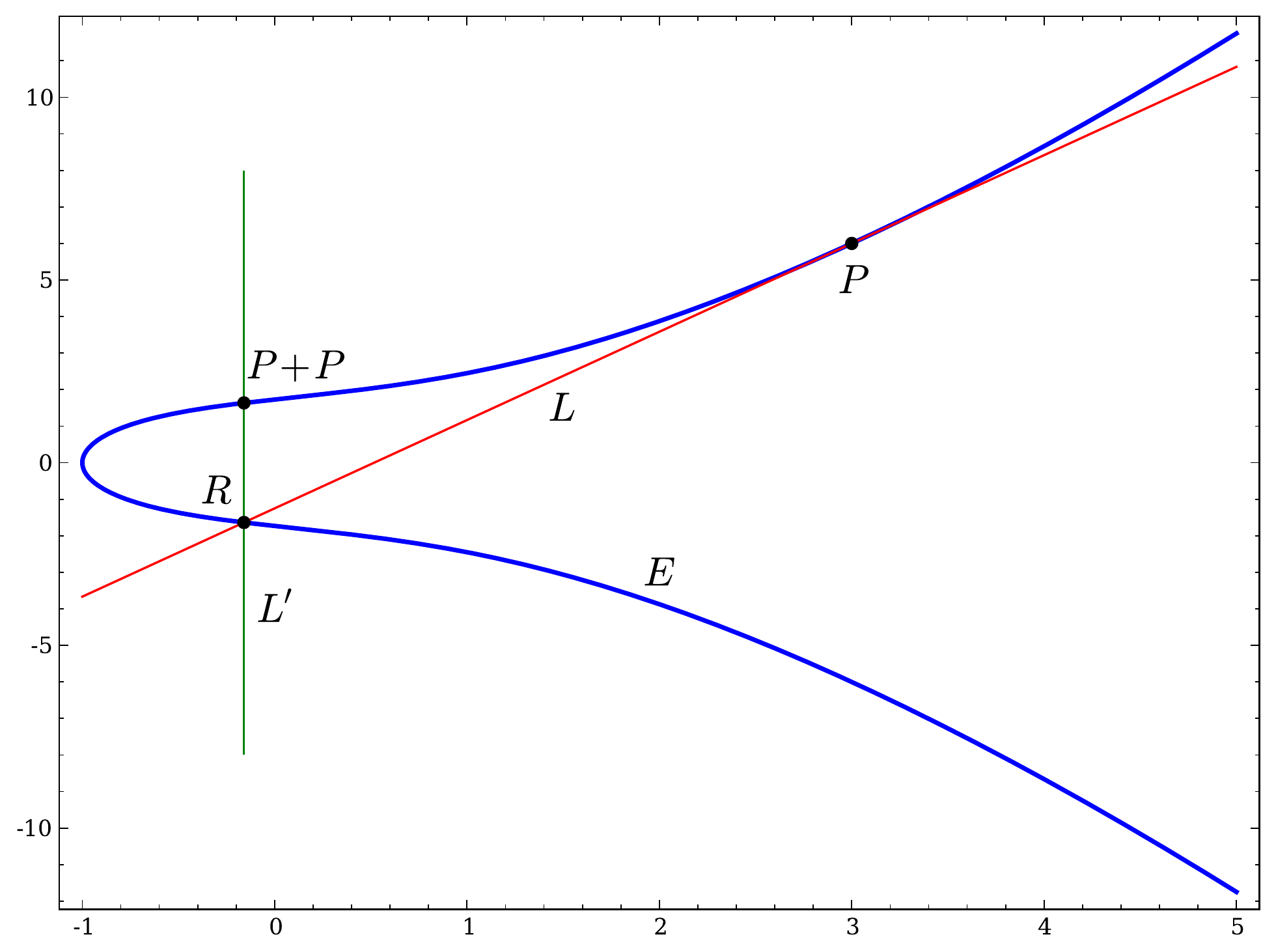}
  \caption[Group law on a cubic]{Example of the group law on the cubic
    $y^2 = x^3 + 2x + 3$; The point $P=(3:6:1)$ is, up to sign, the
    single generator of the Mordell-Weil group $E(\Q)\simeq \Z$. The
    graphics shows how to compute $P+P$.}
  \label{fig:group_law}
\end{figure}
\begin{equation}
  E:\ y^2 = x^3 + f x + g,
\end{equation}
or, by homogenizing, the cubic curve in $\mathbb{P}^2$ given by
\begin{equation}
  E:\ y^2 z = x^3 + f x^2 z + g z^3
  .
\end{equation}
Here, $(x:y:z)$ are homogeneous coordinates of $\mathbb{P}^2$ and the
point at infinity is $O = (0:1:0)$. Given two rational points
$P=(p_x:p_y:p_z)$ and $Q=(q_x:q_y:q_z)$, one can now define a group
action as follows. If $P=Q$, we take $L$ to be the line tangent to
$P$, otherwise we choose it to intersect both points.  By B\'{e}zout's
theorem~\cite{griffiths2011principles}, $L$, being a curve of degree
one, intersects $E$ exactly three times. Therefore, their intersection
defines a third point $R$, which can be shown to be rational
again. This naive map $(P, Q) \mapsto R$ does not form a group law
yet, for example it lacks an identity element. However, this can be
remedied. Take $L'$ to be another line through $R$ and the point at
infinity. Then the third intersection point between $L'$ and $E$
defines the sum $P + Q$, which is now a group action on the set
$E(\Q)$ of rational points on $E$. In particular, $O$ is the identity
element with respect to this action. \autoref{fig:group_law}
illustrates the group law for the single generator of a rank $1$
elliptic curve.

Finally, let us remark that choosing $L'$ to intersect $R$ and $O$ was
an arbitrary choice. Replacing $O$ by any other rational
point $O'$ still gives a valid group law with $O'$
now acting as the identity element with respect to this new group
action.
